\documentclass{aa}

\usepackage{graphicx}
\usepackage{amsmath}
\usepackage{amssymb}
\usepackage{natbib}
\usepackage{color}
\usepackage{enumitem}
\usepackage[T1]{fontenc}

\usepackage{txfonts}
\begin{document}

\title{Planet formation and migration near the silicate sublimation front in protoplanetary disks}




   \author{Mario Flock\inst{1,2} \and  Neal J. Turner\inst{2} \and  Gijs
     D. Mulders\inst{3} \and  Yasuhiro Hasegawa\inst{2} \and  Richard
     P. Nelson\inst{4} \and Bertram Bitsch\inst{1}}

   \institute{Max-Planck Institute for Astronomy (MPIA), K\"{o}nigstuhl 17,
     69117 Heidelberg, Germany \email{mflock@mpia.de}
   \and Jet Propulsion Laboratory, California Institute of Technology, Pasadena, California 91109, USA
   \and Department of the Geophysical Sciences, The University of Chicago, Chicago, IL 60637, USA
   \and Astronomy Unit, Queen Mary University of London, Mile End Road, London E1 4NS, UK}


  \abstract
{The increasing number of newly detected exoplanets at short orbital periods raises questions about
their formation and migration histories. Planet formation and migration depend heavily on the structure and dynamics of protoplanetary disks. A particular puzzle that requires explanation
arises from one of the key results of the Kepler mission, namely the increase in the planetary
occurrence rate with orbital period up to 10 days for F, G, K and M stars.} 
{We investigate the conditions for planet formation and migration near the dust 
sublimation front in protostellar disks around young Sun-like stars. We are especially interested 
in determining the positions where the drift of pebbles would be stopped, and where the migration of 
Earth-like planets and super-Earths would be halted.}
{For this analysis we use iterative 2D radiation hydrostatic disk models which include irradiation by the star, and dust sublimation
  and deposition {depending} on the local temperature and vapor pressure.}
{Our results show the temperature and density structure of a gas and dust
disk around a young Sun-like star. We perform a parameter study by varying the
{magnetized turbulence onset} temperature, the accretion stress, the
dust mass fraction, and the mass accretion rate.
%
{Our models feature a gas-only inner disk, 
a silicate sublimation front and dust rim starting} at around 0.08 au, an ionization
transition zone with a {corresponding} density jump, and a pressure maximum which
acts as a pebble trap at around 0.12 au. {Migration torque maps show Earth- and super-Earth-mass planets halt in our model disks at orbital periods ranging from 10 to 22 days.}} {
{Such periods are in good agreement with both the inferred location of the innermost planets in multiplanetary systems, and the break in planet occurrence 
rates from the Kepler sample at 10 days.} 
In particular, models with small grains depleted produce a trap 
located at a 10-day orbital period, while models with a higher {abundance of} small grains present 
a trap at around a 17-day orbital period.
{The snow line lies at 1.6 au, near where the occurrence rate of the  giant planets  peaks.} We conclude that the
dust sublimation zone is crucial for {forming close-in planets}, especially when considering tightly packed super-Earth systems.}

%


\maketitle

\begin{figure*}[t]
  \resizebox{\hsize}{!}{\includegraphics{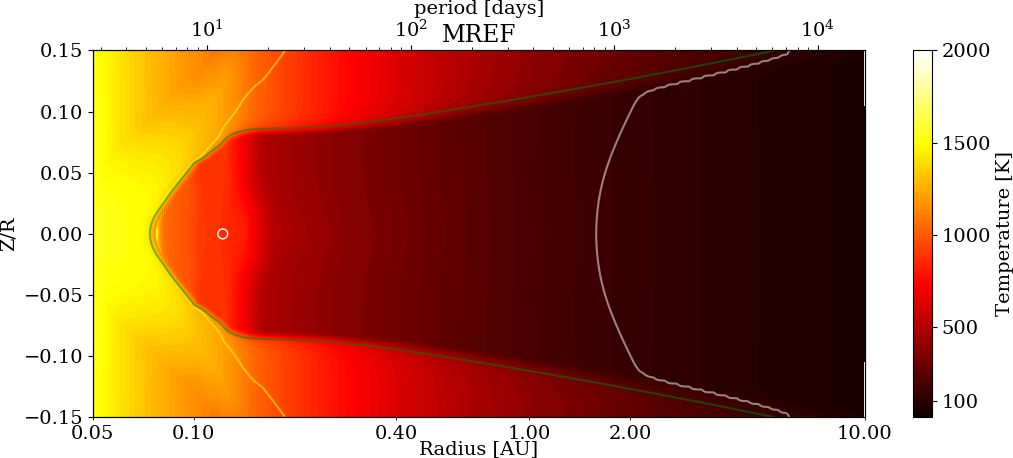}}
\caption{Temperature profile in the 2D {R/Z} plane for our reference model
  \texttt{MREF}, a young solar-type star with a radially-uniform mass
  accretion rate of $\rm 3.6 \times 10^{-9} M_\sun/yr$, a dust-to-gas mass
  ratio for small grains of $0.001$ and stress-to-pressure {ratios} of
  $\alpha_{MRI}=0.01$ and $\alpha_{DZ}=0.001$. The green solid line shows optical depth $\tau = 2/3$ for the starlight irradiation. The yellow dashed
  line shows the silicate {sublimation front}. The white solid line shows the
  water {sublimation front}. The white circle shows the location of the pebble
trap, where {large grains would concentrate}.} 
\label{fig:tempsun}
\end{figure*}

\section{Introduction}
Our {knowledge} of the origins of planetary systems relies on our
understanding of the disks of gas and dust orbiting young
stars. With the growing number of detected exoplanetary systems, we
want to {learn about} their formation and evolution, starting from their 
building blocks in young protoplanetary disks.
One key result of the Kepler mission was the discovery of many planetary systems 
containing Earth-sized and super-Earth planets {orbiting with} periods between 
one day and one year \citep{lis11,bat13,fab14}. 
The Kepler mission focused {on} searching for planetary systems around F-, G-, and K-type stars.
Further analysis of the results showed a clear rise in the occurrence rate
of super-Earths  (rocky planets of several Earth masses) with orbital 
periods longer than 10 days. {The occurrence is uniform at} longer periods \citep{you11,mul15,pet18}.
Over recent years, possible mechanisms for this rise were investigated,
including truncation of the disk by the stellar magnetosphere \citep{lee17}, trapping of
planets at the inner edge of the disk \citep{bra18,car19}, and the removal of
close-in planets by tidal interactions inside 10-day orbital periods \citep{ric12}. Even more recently, \citet{mul18} have found, when including
observational constraints on multiplanetary systems, that on average the  innermost planet
{orbits}  with a 12-day period.

Broadly speaking, models of the formation of short-period planets are based 
either on the drift of pebbles to locations near the inner edge of the disk where they 
concentrate and grow into planets \citep{bol14,cha14} or on planets that form farther  out in the disk, 
perhaps by pebble accretion \citep{lam19,izi19}, and then migrate inward before reaching
locations near the inner edge of the disk where migration halts because the
{disk-planet} torques go to zero \citep{ter07,ida10,fau16,izi17}.
To compare {these planet formation} models with observations, the main disk
property used is   the abundance {of solids \citep{daw15,daw16}.
Formation models are {typically} based on simple assumptions about the inner
disk structure. More detailed self-consistent disk profiles of density
and temperature including irradiation and dust sublimation have not yet  been computed.

In this work our aim is to develop irradiated hydrostatic disk models around young 
Sun-like stars. With respect to the Kepler mission results, we want to address
the influence of the disk structure on both pebble drift and planet migration,
and in particular show how the temperature and density structure leads to
locations where drifting pebbles and migrating planets are trapped.

We follow the numerical methods that we   used previously to model the inner
disk around Herbig Ae/Be-type stars   \citep{flo16}. In this work, we {have}
adapted all the parameters to match a protoplanetary disk around an early Sun-like star, which
is less massive and less luminous than Herbig Ae/Be stars, and for which the disks
sustain a  higher mass accretion rate. 

We first investigate the density and thermal structure of the disk, and
its impact on pebble drift and planet migration. We consider steady-state
models with a radially uniform mass accretion rate. To determine the surface
density we use the temperature dependent $\alpha$ value, which we {directly
calculate} from
our previous 3D radiation magneto-hydrodynamic (MHD) simulations \citep{flo17}. In our previous
work we found that the effect of viscous heating is small \citep{flo16}, and therefore 
we focus on passively heated disks in this work. 

For temperatures below 850-1000K (which is dependent on the density), the ionization level drops
\citep{des15} and turbulent activity due to the magneto-rotational instability (MRI) quickly decreases, 
mainly due to Ohmic resistivity \citep{thi18}. Among the main processes governing the 
temperatures in the region close to the star are the irradiation heating and sublimation of silicate grains,
which change the opacity by orders of magnitude \citep{pol94}, and which we include
in these models.

The structure of the paper is as follows. In Section 2 we briefly review the
main parts of the numerical method. In Section 3 we present the 2D radiation hydrostatic solutions. In
Section 4 we calculate migration maps. Sections 5 and 6 follow with the discussion and conclusions.


\begin{figure*}[ht]
  \resizebox{\hsize}{!}{\includegraphics{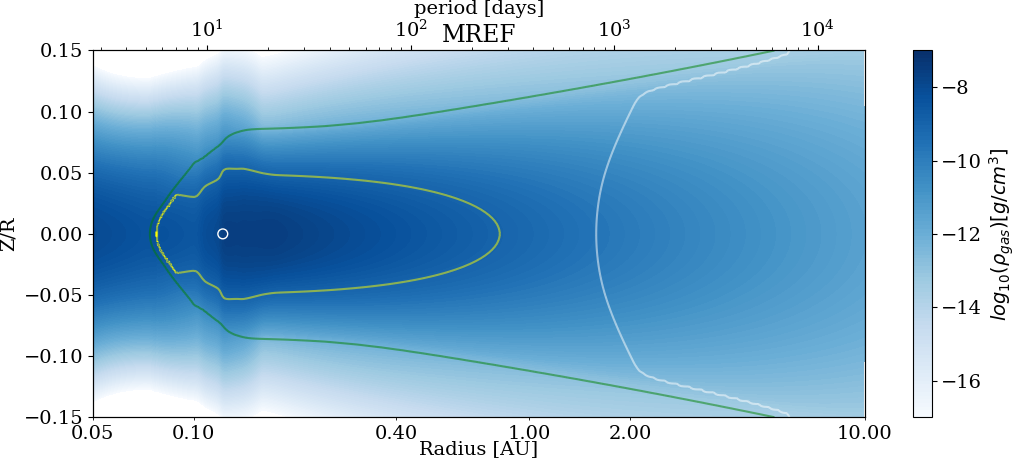}}
\caption{Density profile in the 2D plane of model \texttt{MREF}, a young solar-type star with a
  radially uniform mass accretion rate of $\rm 3.6 \times 10^{-9} M_\sun/yr$, a dust-to-gas mass
  ratio for small grains of $0.001$ and stress-to-pressure ratios of
  $\alpha_{MRI}=0.1$ and $\alpha_{DZ}=0.001$. {Contours of the gas and dust density are shown
  for $10^{-9}$ and $10^{-12} \rm g/cm^3$ (yellow solid line).} The green solid line shows the
  optical depth line  $\tau = 2/3$ for the starlight irradiation. The white solid line
  shows the water snow line in the disk. The white circle shows the location of the pebble
trap.} 
\label{fig:denssun}
\end{figure*}

\section{Method}
We followed the previous numerical method presented by \citet{flo16}.
We first calculated a radiation hydrostatic equilibrium, solving iteratively
the radiation transfer including irradiation. In the first step the gas surface density profile was determined using
\begin{equation}
\rm \Sigma(R) = \frac{\dot{M}}{3 \pi \nu_t(R)}
\label{eq:sig_mdot}
\end{equation}
at the cylindrical radius $\rm R$, assuming a constant radial mass accretion
rate $\rm \dot{M}$. The viscosity  is given by
\begin{equation}
\rm \nu_t = \frac{\alpha c_s^2}{\Omega} \, ,
\label{eq:nu_t}
\end{equation}
where $\rm c_s$ is the sound speed, $\rm \Omega=\sqrt{GM_*/R^3}$ is the disk rotation frequency,
and the stress-to-pressure ratio is given by
\begin{equation}
\rm \alpha = (\alpha_{MRI} - \alpha_{DZ} )  \left [ \frac{1-\tanh{\left(
    \frac{T_{MRI}-T}{25 K} \right )}} {2} \right ] + \alpha_{DZ} \, ,
\label{eq:alpha}
\end{equation}
with $\alpha_{MRI}=0.01$ being the stress-to-pressure ratio in the active zone
$(T>T_{MRI})$, $\alpha_{DZ}$ being the stress-to-pressure ratio in the dead zone, and
$T_{MRI}$ being the ionization transition temperature for the MRI to operate.

The initial temperature field $\rm T(r,\theta)$ was calculated using the
optically thin solution, {which depends on the stellar luminosity and the opacities}.
We then calculated {the density} $\rm \rho(r,\theta)$ and {azimuthal
velocity} $\rm v_{\phi}(r,\theta)$ by solving {the governing equations in
spherical coordinates $\rm (r,\theta,\phi)$ with axisymmetry to obtain the solution} for hydrostatic equilibrium,
\begin{eqnarray}
\rm \frac{\partial \mathrm{P}}{\partial \mathrm{r}} &=& - \rho \frac{\partial \Phi}{\partial
\rm  r} + \frac{\rho \rm v^2_\phi}{\rm r} \label{eq:P_R} \\
\rm \frac{1}{r} \frac{\partial \mathrm{P}}{\partial \theta} &=&
\rm \frac{1}{\tan{\theta}}\frac{\rho v^2_\phi}{r} \label{eq:P_T} \, ,
\end{eqnarray}
where $\rm \Phi$ is the gravitational potential $\rm \Phi = GM_*/r$ (with $\rm G$ being the gravitational constant and $\rm M_*$ 
  the stellar mass) and $\rm P$ is the thermal pressure that relates to the
temperature through the ideal-gas equation of state:
\begin{equation}
\rm P= \frac{\rho k_B T}{\mu_g u}
\end{equation}
with the mean molecular weight $\rm \mu_g$, the Boltzmann constant $\rm k_B$, and the atomic mass unit $\rm u$.

For a given density field, the {internal energy and radiation field in
  radiative} equilibrium were obtained as the steady-state solution to the
following coupled pair of equations {describing heating, cooling, and the
  flux-limited diffusion of the disk's thermal radiation}:
\begin{align}
\label{eq:RAD1}
\begin{split}
\rm \frac{1}{\gamma-1}\partial_t P &= \rm - \kappa_P \rho \mathrm{c} (a_R T^4 - E_R) - \nabla \cdot F_*, \\
\rm \partial_t E_R - \nabla \frac{c \lambda}{\kappa_R \rho} \nabla
E_R &= \rm + \kappa_P \rho \mathrm{c} (\rm a_R T^4-E_R). 
\end{split}
\end{align}
Here   we used the adiabatic index   $\rm \gamma$, the radiation energy E$_\mathrm{R}$,
the irradiation flux $\mathrm{F}_*$, the flux limiter $\rm \lambda$
\citep{lev81}, {the Rosseland and Planck mean opacity $\rm \kappa_R$ and
  $\rm \kappa_P$,} the radiation constant $\rm a_R=4 \sigma_b/c$ (with the
Stefan-Boltzmann constant $\rm \sigma_b$ and    the speed of light c). {We computed $\mathrm{F}_*$ by solving the transfer equation along radial rays
from the central star, treated as a point source.} {Following our previous
work, we simplified the problem by assuming $\rm \kappa_R =\kappa_P$ and considering three opacities: the gas opacity $\rm
  \kappa_{gas}$, the  Planck dust opacity at the dust sublimation temperature $\rm
  \kappa_P(T_{s})$, and the Planck dust opacity at the stellar temperature
  $\rm \kappa_P(T_*)$.}

{We took the gas opacities from \citet{mal14} for  starlight and
for the disk's thermal emission.  The two dust opacities were frequency-averaged,
one over the blackbody spectrum at the sublimation temperature and the other
over the irradiating starlight spectrum.  The combined opacity was determined
using the dust-to-gas ratio.} For details on the gas and dust opacities we refer to
the Appendix.

{We ignored the heat released by accretion within the disk.} Viscosity was only used to set the
constant mass accretion and to obtain {a surface density profile
  consistent with the uniform mass accretion rate.} 
In our previous work we found the effect  of viscous heating remains small \citep{flo16}, 
while   recent disk wind models predict an even smaller viscous heating effect 
\citep{bet17,mor19} in the dead-zone region. We   address this again in the discussion section.

Convergence was reached by iteratively solving for the radiation equilibrium
and hydrostatic equilibrium, including irradiation from the star.
We iterated until we reached convergence. We iterated 30 times (compared to 20
times in our previous method) {as these models span a larger domain}. 

\subsection{Dust sublimation}
\label{sec:dust_sub}

As in our previous work, we followed \citet{pol94} and used the fitting model of
\citet{ise05} that applies to situations for which the most refractory
grains are silicates when determining the {sublimation} temperature:
\begin{equation}
\rm T_{s}=2000K \left ( \frac{\rho}{1 g\, cm^{-3}} \right )^{0.0195}
\, . 
\label{eq:ev}
\end{equation}
The value of $\rm T_{s}$ was then used to calculate the dust-to-gas ratio $\rm f_{D2G}$,
which is  the  ratio between the dust density $\rm \rho_d$ and the gas
density. 
We slightly modified our previous formula (Eq. 11 in \citealt{flo16}) to
\begin{equation}
\rm f_{D2G}=\left\{
                \begin{array}{l}
                  \rm f_{\Delta \tau} \left\{ \frac{1-\tanh( \left (
                  \frac{T-T_{s}}{100K} \right )^3)}{2} \right\} \left\{
                  \frac{1-\tanh(2/3-\tau_*)}{2} \right\} \\
                  \rm f_0 \, \rm for\, T<T_{s} \rm \, and \, \tau > 3.0  
                \end{array} \right.
\label{eq:d2g}
,\end{equation}
with the {additional constraint that $\rm f_{D2G} \le f_0$, with $\rm f_0$
  being the} maximum dust-to-gas mass ratio and $\rm f_{\Delta
  \tau}=0.2/(\rho_{dust} \Delta r \kappa_P(T_*)$ being the dust amount to
account for optical depth of $\Delta \tau =0.2$ {for a given grid cell with
radial width of $\rm \Delta r$}. This value ensures that we resolve the absorption of the incoming radiation at the
inner rim. We slightly {steepened the transition} compared to our previous work where
we used $\Delta \tau =0.3$. We also {centered the absorption} at $\tau=2/3$ {where} most of the absorption happens at $\tau=2/3$.

For $\rm T<T_{s}$ and $\tau > 3.0$ most of the irradiation is absorbed and the
available {silicates occur} in solid form.
For more information on the method, the
flux limiter, and the treatment of the optical depth at the rim wall we refer
to \citet{flo16}.

\section{Model parameters}
The model input parameters are summarized in table~\ref{tab:info} and table~\ref{tab:model}.
\begin{table}
\begin{tabular}{lll}
\hline
\hline
Stellar parameters & $\rm T_*=4300\, K$, $\rm R_*=2.6\, R_\sun$\\ 
                  & $\rm M_*=1.0\, M_\sun$\\
Opacity & $\rm \kappa_P(T_*)=1300\, cm^2/g$\\
        & $\rm \kappa_P(T_{s}) = 700\, cm^2/g$\\ 
        & $\rm \kappa_{gas} \sim 10^{-5}\, cm^2/g$\\
Grid size $\rm N_r \times N_\theta $ & 2048 x 128 \\ 
Cell aspect ratio & $ \Delta R / R \Delta \theta \sim 1.1$ \\
Radial domain $\rm R_{in}-R_{out}$ & 0.05-10 AU \\
Vertical domain $\rm Z/R $ & $\pi/2 \pm 0.15$\\ 
Ionization transition & $\rm T_{MRI} = 900K$\\
  $\alpha$ viscosity &  $\alpha_{MRI} = 10^{-2}$ {\tiny for $T > T_{MRI}$}\\
                       & $\alpha_{DZ} = 10^{-3}$ {\tiny for $T \leq T_{MRI}$}\\
  Dust-to-gas mass ratio & $\rm f_0=10^{-3}$\\
  of small grains {\small ($\le 10 \rm \mu m$)}\\
Mass accretion rate        & $\rm \dot{M}=3.6 \times 10^{-9} M_\sun/yr$\\
\hline
\hline
\end{tabular}
\caption{Parameters for the \texttt{MREF} model.}
\label{tab:info}
\end{table}
The stellar parameters were taken from a stellar evolution
track by \citet{sie02} and \citet{dan94}, and  represent a
young solar star with an age of approximately one million years. Here we recalculated the stellar
spectrum-weighted irradiation opacity. The {redder} spectrum of the T Tauri
star reduces the irradiation opacity compared to our previous models, which
were calculated for a Herbig-type star. {The irradiation opacity $\rm \kappa_P(T_*)$ is 1300 $\rm cm^2/g$ and the thermal emission opacity is 
$\rm \kappa_P(T_{s})=700\, cm^2/g$.}

The dust-to-gas mass ratio {ranges from} $\rm 10^{-4}$ to $10^{-3}$
representing grains with sizes up to 10 $\rm \mu m $, {which} dominate the opacity at the relevant temperatures and
wavelengths. {Modeling indicates that most of the solid material is locked in larger grains
\citep{bir12}, consistent with a protostellar disk  millimeter flux-radius
correlation \citep{ros19} and the weakness of silicate features in the infrared spectra of the disks around
young stars with masses near solar \citep{fur06,fur09,fur11}.}

{We investigated ionization transition temperatures of} 900K and 1000K. For our reference model we assume 900K, motivated by   the ionization thresholds, which {are} at cooler temperatures for higher
densities \citep{des15}. {We studied accretion rates} between $10^{-9}$ and
$\rm 10^{-8} M_\sun/yr$ corresponding to {those of} disks in the Chamaeleon I
star-forming region \citep{man17}. For our reference model, \texttt{MREF}, we
used a value of $\rm 3.6 \times 10^{-9}
M_\sun/yr$. We note that as we did not include the outer edge of the
disk in our models, the total disk mass remains a free parameter.

For the {stress-to-pressure ratio} $\alpha_{MRI}$ in the MRI active region, we took values between $0.01$ and $0.1$.
The larger value was determined from our previous 3D
radiation non-ideal MHD simulations. In \citet{flo17} we found a
{stress-to-pressure ratio} of up to 0.1 for the net vertical magnetic flux case. For temperatures
below the ionization {threshold} we chose {stress-to-pressure ratios
  $\alpha_{DZ}$} between $10^{-3}$ {and} $5 \times 10^{-4},$ to mimic the accretion activity either by
hydrodynamical instabilities \citep{lyr19} or a magnetically driven wind
\citep{bet17}. We note again that the increase in the stress-to-pressure ratio
at the ionization transition has a direct impact on the surface density
profile. Compared to our previous models \citep{flo16} we also investigated
larger jumps in $\alpha$ and the surface density.

{The grid resolution is} 2048 cells in radius
{logarithmically spaced} and 128 in the meridional direction for a domain extending from 0.05 to 10 AU in
radius and {to elevation angles 0.15 radians on either side of the equatorial
plane} in the meridional direction. 
The inner radial boundary was chosen to be
close to but still {outside the radius where the stellar magnetic field is expected to truncate the disk.}

\begin{figure}
  \resizebox{\hsize}{!}{\includegraphics{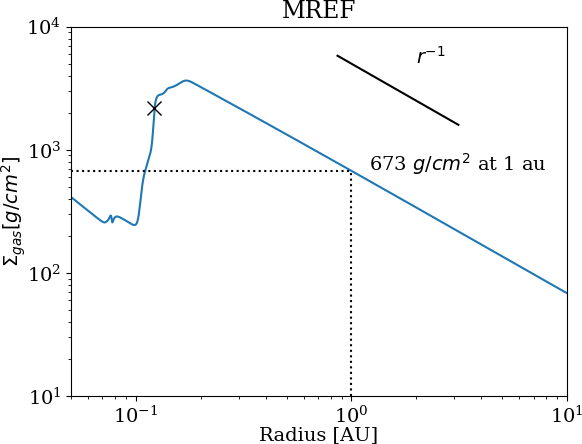}}
  \resizebox{\hsize}{!}{\includegraphics{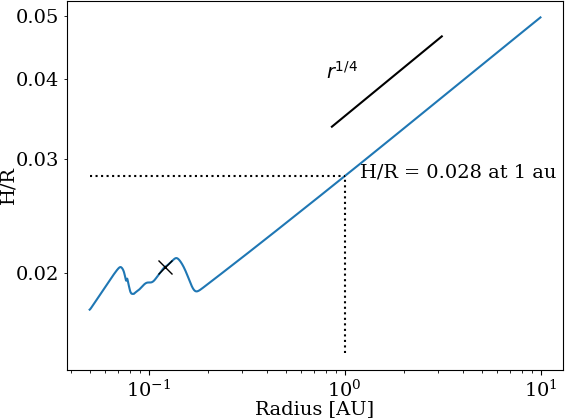}}
  \caption{Top: Radial profile of the gas surface density for our reference
    model \texttt{MREF}. The steep rise at
  around 0.11 au is due to the drop in the accretion stress. Bottom:
  Profile of the {midplane} scale height H/R. {The markings in the panels
  show} the values at 1 au (dotted lines), the position
of the pebble trap (black cross, $\times$) and typical slope profiles for comparison (black lines).} 
\label{fig:surf0}
\end{figure}

\begin{figure}
  \resizebox{\hsize}{!}{\includegraphics{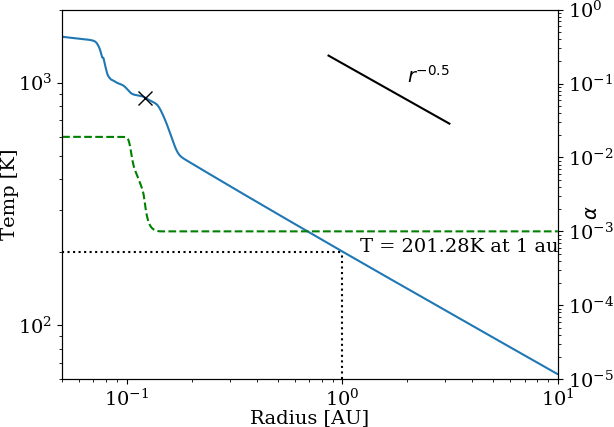}}
  \caption{Radial profile of the disk {midplane} temperature. Shown in the plot are the values at 1 au and the position
  of the pebble trap (black cross, $\times$). The {midplane stress-to-pressure} ratio
  $\alpha$ is {overplotted} (green dashed line).} 
\label{fig:temp0}
\end{figure}

\section{Results}
The results derived from the radiation hydrostatic solution for our
reference model are shown in Fig.~\ref{fig:tempsun} and
Fig.~\ref{fig:denssun}. Figure~\ref{fig:tempsun} presents the 2D temperature profile 
of the disk, showing a similar structure to that found for a Herbig-type
star \citep{flo16}, but where the key features occur on smaller radial scales closer to the star:
a hot dust halo in front of the rim inside 0.08 au, a curved dust rim between 0.08 and 0.15 au, 
a small shadowed region between 0.2 and 0.3 au, and a flared disk beyond 0.3
au. The {starlight optical depth} $\tau=2/3$ line starts at the midplane
at 0.08 au, while reaching an {elevation angle} of $\rm \theta=0.08 rad$ at 0.15 au. At 8 au the $\tau=2/3$ line reaches $\theta=0.15$,
which corresponds to roughly 3 scale heights. 
We {find} a pressure maximum and pebble trap located at 0.12 au, indicated by a white circle in Fig.~\ref{fig:tempsun}. In this
region, inward drifting {pebbles} are trapped, and could possibly trigger planetesimal formation. 
Figure~\ref{fig:denssun} presents the corresponding 2D gas density profile.
{The next feature with increasing distance from the star is} the rise in gas density at 0.13 au due to the
lower stress-to-pressure ratio at temperatures below the ionization threshold. 
Midplane densities of $\rm 10^{-8}\, g cm^{-3}$ are reached in the
gas. The density jump is clear {in the gas surface density profile} in the top panel of
Fig.~\ref{fig:surf0}. For this model we
obtain a gas surface density of 680~$\rm g/cm^2$ at 1 au. Figures~\ref{fig:surf0} and
 ~\ref{fig:temp0} show the radial profile of the disk scale-height--to--radius ratio ($H/R$) and the {midplane} temperature. {Near the silicate sublimation
  front the} $H/R$ is around 0.02, {generally increasing outward  to reach} $0.028$ at 1 au. The disk radial
temperature shows the typical profile, with a shallower decrease at the rim at
around 1000~K, while at 1 au the temperature reaches roughly 200~K. {For comparison, Fig.~\ref{fig:surf0} and Fig.~\ref{fig:temp0} also show  the
  slopes for $r^{-1}$ for density, $r^{2/7}$ for $H/R$, and $r^{-0.5}$ for the
  temperature.}

Figures~\ref{fig:tempsun} and  ~\ref{fig:denssun} both include the snow line.
{We locate the snow line using the water vapor mole fraction}
\begin{equation}
    X_{H_2O}= \left \{
      \begin{array}{ll}
        1.2 \times 10^{-3} \,\, (1.2 \times 10^{-3} P \le P_{sat} (T) ),\\
        P_{sat}(T)/P  \,\, (1.2 \times 10^{-3} P \ge P_{sat} (T) ),\\
      \end{array}
      \right .
\end{equation}
\citep{oka11} {with a total water mole fraction of $1.2 \times 10^{-3}$ and a} saturated vapor pressure
\begin{equation}
P_{sat} = e^{-6070 K/T+30.86} dyn \, cm^{-2}.
\end{equation}
In Fig.~\ref{fig:tempsun} we determine the position where half of the water is in the form of ice, 
such that $x_{ice}=0.5$, using
\begin{equation}
x_{ice}= 1 - X_{H_2O}/(1.2 \times 10^{-3}).
\end{equation}
Using this formula we determine that the snow line crosses the midplane at 1.6 au,
reaching 2 au at 2 scale heights  above the midplane. Due to the lower pressure
and higher temperature, the snow line moves radially outward with increasing
vertical height until it reaches 7 au at 3 scale heights.
The snow line is an important location for planet formation. {In
  particular, due to the} predicted agglomeration of
larger grains {this is a preferred location of giant planet formation}
\citep{oka11,ros13,dra17,sch18}. {We note that for our choice of stellar
  parameters and resulting stellar luminosity of $\sim 1.92 L_{\sun}$ the snow
  line is located farther out radially  compared to models with a lower stellar
  luminosity. This value is similar to recent solar-mass and metallicity
  stellar models \citep{ama19}, which predict roughly 2 $L_\sun$ for a star
  one million years in age.}
\begin{figure*}[ht]
\resizebox{\hsize}{!}{\includegraphics{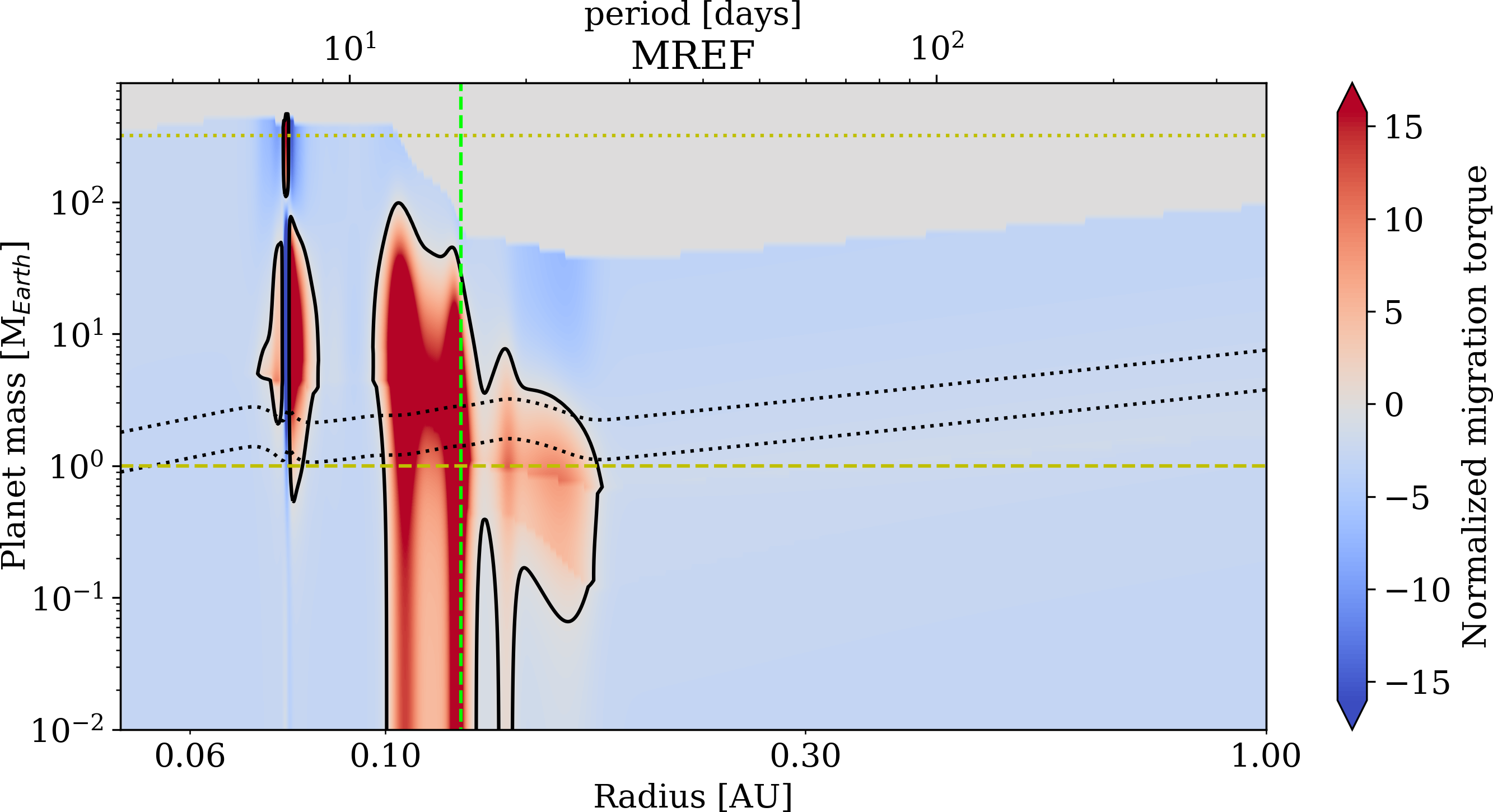}}
\caption{Normalized total {orbital migration} torque ($\Gamma_{\rm tot}/\Gamma_0$) 
  vs. planetary mass and orbital radius in the reference model \texttt{MREF}, {zoomed into
  the region inside 1 au}. Inward migration is shown in blue,
   while  outward migration is shown in red. The solid black  line marks
  where the migration torque is zero, and hence where planets are trapped. Dotted lines indicate the pebble isolation mass, following
  \citet{lam14} (lower) and \citet{bit18} (upper). The green dashed line indicates the location of
  the pebble trap, where the inward radial drift of solid particles halts. {The gray shading at the top} shows the gap-opening region {according to} \citet{cri06}. 
  Here, planets would modify the surface density structure strongly. The {horizontal} yellow 
  dashed and dotted lines indicate the {masses of} Earth and Jupiter.} 
\vspace{8pt}
\label{fig:torq}
\end{figure*}

\begin{table}[ht]
\begin{tabular}{p{0.5cm}p{0.5cm}p{0.5cm}p{1.0cm}p{1.5cm}p{0.5cm}p{0.5cm}p{0.5cm}}
  Model & $T_{MRI}$ &  $\alpha_{MRI}$ & $\alpha_{DZ}$ &  $\dot{M} $ & $f_0$
  & $\kappa_{gas} $ & trap\\
\hline
{\tiny MREF} & 900 & 0.01 & $10^{-3}$ & $3.6 \cdot 10^{-9} $ & $10^{-3}$ & $10^{-5} $ & 0.12\\
{\tiny M0}   & 900 & 0.05 & $10^{-3}$ & $3.6 \cdot 10^{-9} $ & $10^{-3}$ & $10^{-5} $ & 0.13\\
{\tiny M1}   & 900 & 0.05 & $10^{-3}$ & $3.6 \cdot 10^{-9} $ & $10^{-4}$ & $10^{-5} $ & 0.093\\
{\tiny M2}   & 900 & 0.1 & $10^{-3}$ & $3.6 \cdot 10^{-9} $ & $10^{-3}$ & $10^{-5} $ & 0.13\\
{\tiny M3}   & 900 & 0.05 & $5\cdot 10^{-4}$ & $3.6 \cdot 10^{-9} $ & $10^{-3}$ & $10^{-5} $ & 0.15\\
{\tiny M4} & 900 & 0.05 & $10^{-3}$ & $10^{-8} $ & $10^{-3}$ & $10^{-5}$ & 0.15\\
{\tiny M5} & 900 & 0.05 & $10^{-3}$ & $10^{-9} $ & $10^{-3}$ & $10^{-5}$ & 0.13\\
{\tiny M6} & $10^3$ & 0.05 & $10^{-3}$ & $3.6 \cdot 10^{-9} $ & $10^{-3}$ & $10^{-5}$ & 0.12\\
{\tiny M7} & 900 & 0.05 & $10^{-3}$ & $3.6 \cdot 10^{-9} $ & $10^{-3}$ & $10^{-4}$ & 0.13\\
\hline
\end{tabular}
\caption{Model name, ionization transition in Kelvin, MRI $\alpha$ value, $\alpha$ in
  the dead zone, mass accretion rate in $[M_\sun/yr]$, dust-to-gas mass ratio,
  gas opacity in $\rm [cm^2/g]$, and resulting
  location of the pebble trap in au.}
\label{tab:model}
\end{table}

\subsection{Migration maps}
In this section we calculate the migration torques on planets embedded in our
reference model. We are especially interested in determining where {orbital migration halts, as this may set up the architecture of the new planetary system.}

{We use} the torque {formula from} \citet{paa11} as adapted by
\citet{bit11}. The total torque $\Gamma_{tot}$ is the sum of the Lindblad
torque $\Gamma_L$ and the corotation torque $\Gamma_c$. The Lindblad torque
arises because of the spiral waves launched at {the planet's} Lindblad resonances. We follow \citet{paa08} and write 
\begin{equation}   
\Gamma_L=\frac{\Gamma_0}{\gamma} \left ( -2.5 -1.7 \beta^s + 0.1 \alpha^s \right )
\end{equation}   
with $\alpha^s$ the negative slope of the surface density profile $\Sigma
\propto r^{-\alpha^s}$ and $\beta^s$ the negative slope of the temperature profile $T \propto r^{-\beta^s}$. The normalization torque $\Gamma_0$ is defined as 
\begin{equation}   
\Gamma_0=\left ( \frac{q}{h} \right )^2 \Sigma_P r_P^4 \Omega_P^2 
\end{equation}   
with $q$ the planet-to-star mass ratio, $h$ the aspect ratio $h=H/R$,
$\Sigma_P$ the surface density at the planet's location $r_P$, and $\Omega_P$ the
orbital frequency of the planet. Calculating the corotation torque $\Gamma_c$
is more complicated because this can saturate at a level depending on planet
mass, disk viscosity, and local thermal diffusion rate.
We {compute $\Gamma_c$ as detailed in} Sects. 5.6 and 5.7 of \citet{paa11}.

The above calculation of $\Gamma_{tot}$ is valid until the
gap-opening or type II migration regime is reached. To determine the gap-opening mass
we follow the equation by \citet{cri06}:
\begin{equation}   
\frac{3}{4} \frac{H}{R_H} + \frac{50}{q Re} = 1.0 
\label{eq:gap}
,\end{equation}   
with the Hill radius
\begin{equation}   
R_H=R_p \left( \frac{q}{3} \right )^{1/3}
\end{equation}   
and the Reynolds number
\begin{equation}   
Re=R_p^2 \Omega_p /\nu_t.
\end{equation}
{Planets exceeding this mass} would substantially change the surface density structure,
and hence the migration torque.

In Fig.~\ref{fig:torq} we present a map of the total torque in the planet orbital radius-mass plane. For this calculation, the values of the density,
temperature, stress-to-pressure ratio, surface density, and opacity are all taken from the reference radiation
hydrostatic solution which is in inflow equilibrium. The figure shows that {across} most of the inner disk and for
most planetary masses, {migration is} inward,   as expected from type I migration models
\citep{war97,fog07,cri17}. {From left to right, the first
  region of outward migration   is connected to the dust sublimation zone at around 0.08 au and
  the resulting drop in temperature. This region is relevant only for planets
  more massive than the Earth and it could be a halting point for planets all
  the way up to the gap-opening mass.

The next and most prominent region of outward migration lies in the steep rise of the surface density starting at 0.1 au. Here outward
migration occurs for planets with masses from as low as 1\% of an Earth
mass and up to 100 Earth masses. The normalized
migration torque is very strong, reaching $\Gamma_{\rm tot}/\Gamma_0 \ge 15$. Such a planet trap
at surface density transition was already proposed by \citet{mas06}. The third
outward migration region is due to the small shadowed zone where the
temperature drops significantly starting at 0.13 au. This outward migration region affects planets with masses between 0.1 and 4 Earths.} The radii where the torque vanishes (black solid contour lines) are 
especially interesting for planet formation and
evolution models. At these locations, the Lindblad and corotation
torques exactly cancel each other.

In Fig.~\ref{fig:torq} we overplot with a vertical green dashed line
the location of the pressure maximum where inward-drifting pebbles
accumulate. This location {approximately coincides with} the planet migration trap.
If a steady radial drift of pebbles is present, we might expect a first
planetary core to form at this location. Growth by {accreting} pebbles could continue
until the planet reaches the pebble isolation mass. \citet{lam14} found the pebble isolation mass to be
\begin{equation}                                                                                               
M_{iso}=20 M_{Earth} \left ( \frac{h}{0.05} \right )^{3}.                                                    \end{equation}
Recently, \citet{bit18} found that the pebble isolation mass can, for certain
parameters, be a factor of two higher than the fit by \citet{lam14}. 
In Fig.~\ref{fig:torq} we show both estimates.   
{From this calculation we expect the planets to form at the
pebble trap near 0.13 au and to have} masses between 1.5 and 3
Earth masses. 

Finally, {gray shading} indicates the region of type II migration
where the planet's {tides would be strong enough to} open a gap. {Such
a drastic change in the disk structure means the migration torque calculation
is invalid in this region.} {We also note that  we do not see any change
  in the migration zones radially outward between 1
and 10 au. In Sect. 4.3} we   compare these results with the exoplanet occurrence rate
obtained from the Kepler sample.
\subsection{Parameter study}
In this section we vary {the} parameters (see Table~\ref{tab:model}). A summary of {the resulting} temperature and surface density profiles
is in Fig.~\ref{fig:partemp}. Increasing the value of $\alpha_{MRI}$ in the
inner disk in model \texttt{M0} compared to \texttt{MREF} leads to a steeper
gradient in the surface density. This also {affects} the
planet trap, {extending} the zero-torque region to larger planet masses (see Appendix A).
The position of the pebble trap is shifted
outward by only 0.01 au. Decreasing the abundance of small grains, and hence the opacity, has
a larger effect. Decreasing the dust-to-gas ratio to $10^{-4}$ in
model \texttt{M1} shifts the inner rim closer to the star, and results in a steeper rise
in the temperature at the rim. The pebble trap moves radially inward to 0.093 au, corresponding
to a 10-day orbital period. The other noticeably different model is \texttt{M4} with the highest abundance of small grains, which shifts the pebble
trap to 0.15 au, corresponding to a 22-day orbital period. 
Increasing the gas opacity by one order of magnitude  in model
\texttt{M7} had only a small effect on the structure as the gas between
the sublimation front and the star remains optically thin. 

\begin{figure}
  \resizebox{\hsize}{!}{\includegraphics{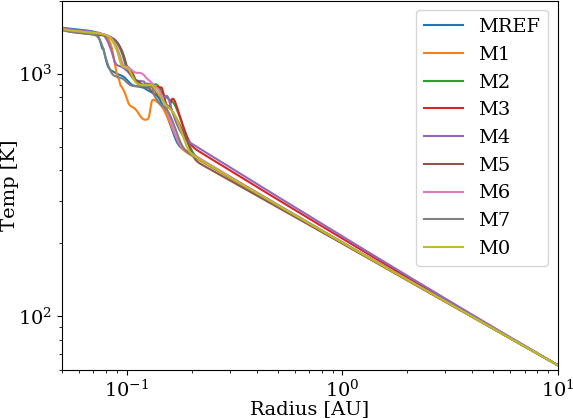}}
\resizebox{\hsize}{!}{\includegraphics{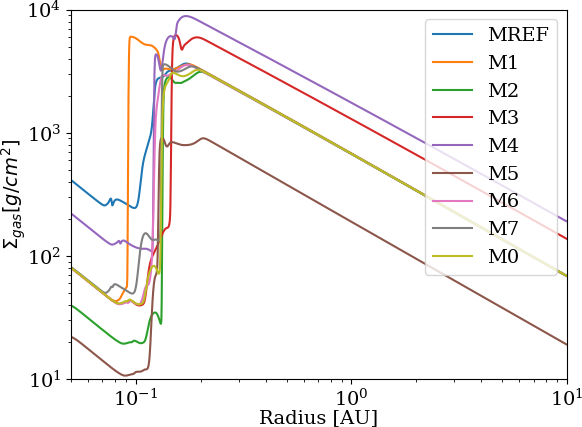}}
\caption{Temperature and surface density profiles over radius for all models.} 
\label{fig:partemp}
\end{figure}

\subsection{Comparison with exoplanet occurrence rates}
In this section we compare our results with previous analyses of
exoplanet occurrence rates. We note that   for all models we use a fixed
stellar type, while the Kepler sample is based on a mix of F-, G-, and K-type stars.

{In the top panel of  Fig.~\ref{fig:mig}  we show the inferred real distribution of the orbital period of the innermost
planet in Kepler multiplanetary systems reported by \citet{mul18}.}\footnote{
  Here ``inferred real distribution'' means that the results are corrected for  
  observational biases \citep{mul18}.}
{Overplotted are the positions of the pebble} traps we found in our
models.

The pebble trap {in our models} reproduces quite well the position of the
innermost planet in multiplanetary systems. The pebble trap occurs slightly farther from the star, with a mean orbital period of around 17 days.
{The dust-depleted} model, \texttt{M1}, {matches the 10-day peak in
  the measured planets' distribution}. \citet{mul18} show  that {the
  innermost planets had} between one and four Earth radii, roughly corresponding to
between one and ten Earth masses. {Interestingly,} this {is similar to the mass range} of the type
I migration trap (Fig.~\ref{fig:torq}). The scatter in {the  measured} radii (and masses, assuming a an Earth-like density) of the innermost
  planets is consistent with
the {range from} the pebble isolation mass {up to} the gap-opening
{threshold}. 

{The innermost planet trap in Fig.~\ref{fig:torq} is located near an 8-day
  orbital period and connected with the dust sublimation front.} Inward-migrating super-Earths can also pile up at the planet trap at the inner
edge of the gas disk where a cavity forms because of the stellar magnetosphere.
{That inner-edge trap} sits closer to the star than the trap in our
models, which is {a result of} the transition between the inner MRI-active region and the
dead zone. {We further note that multiplanetary systems in resonant chains can push the innermost planets farther in \citep{car18,izi19,car19}}.

In the bottom panel of Fig.~\ref{fig:mig}  we {show} the results of
\citet{fer18} {for the occurrence rate of  giant planets from RV surveys and Kepler. They find a clear peak around 2 au, which is very close to the
predicted position of the snow line in our models.

\begin{figure}
  \resizebox{\hsize}{!}{\includegraphics{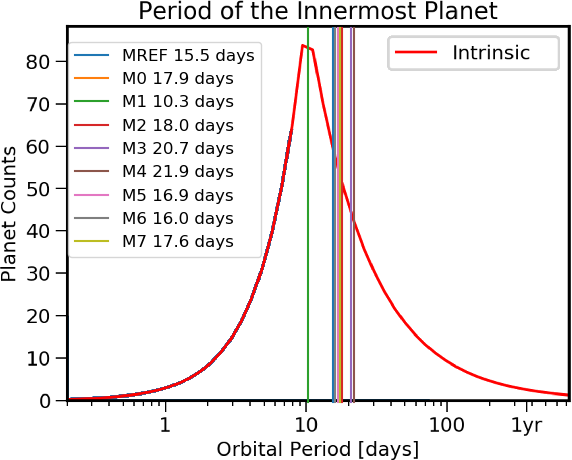}}
  \resizebox{\hsize}{!}{\includegraphics{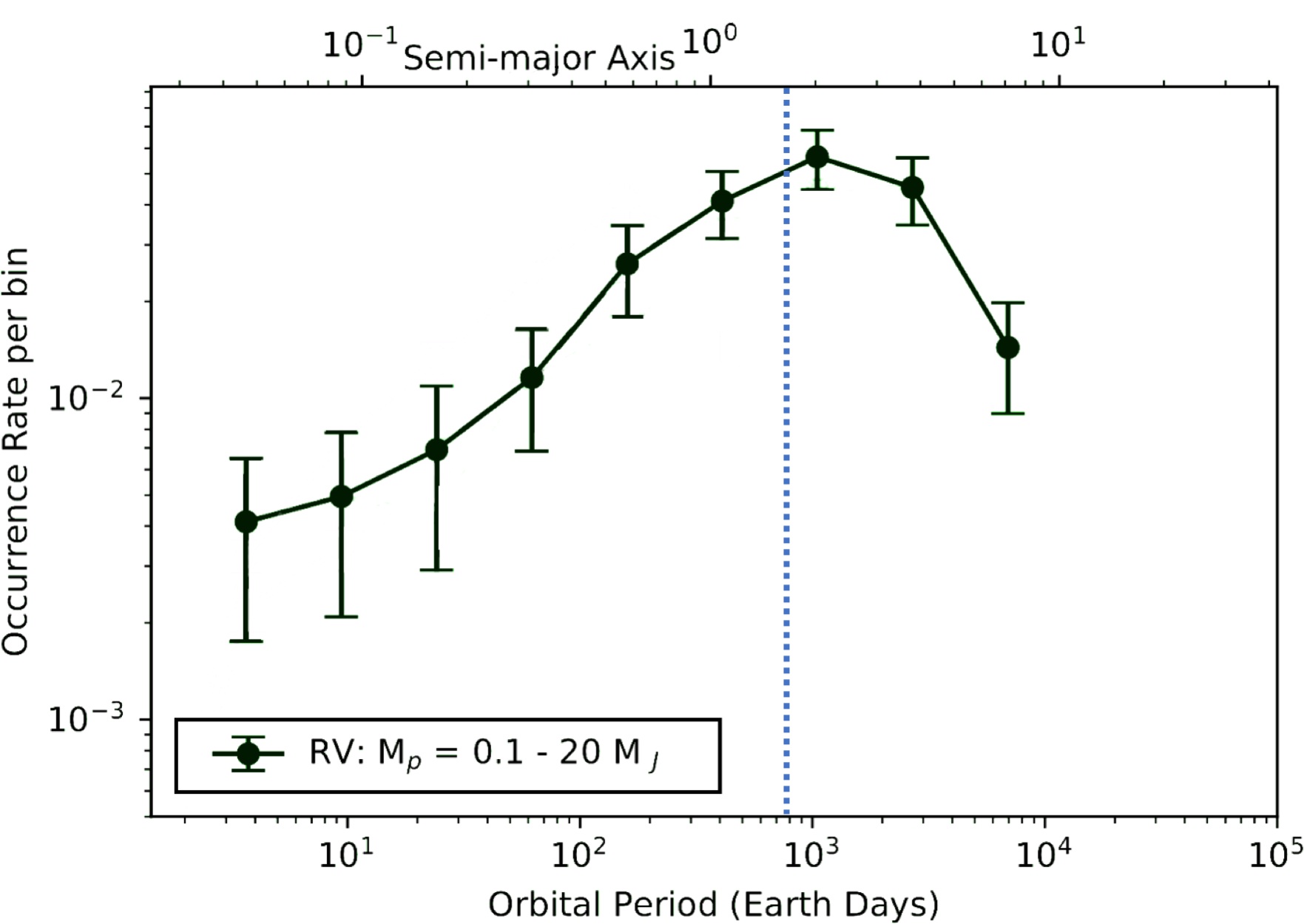}}
  \caption{Top: Orbital period of the innermost planet in Kepler multiplanetary
    systems, taken from \citet{mul18}. Overplotted are the positions of the
    pressure maxima for our models. Bottom: Giant planet occurrence
  rate for RV and Kepler surveys \citep{fer18}. The {blue dashed}
  line indicates the position of the snow line in our models.} 
\label{fig:mig}
\end{figure}

\subsection{Dependence on luminosity}
The position of the sublimation front
{is determined by} the stellar luminosity \citep{dul10}.
{The position of the pebble trap has a similar dependence, as shown in
  Fig.~\ref{fig:lum} where we combine model \texttt{MREF} with the results for
  higher luminosity stars from \citet{flo16}. The location of the pressure maximum  varies with  stellar luminosity to the power 0.6.}


\begin{figure}
  \resizebox{\hsize}{!}{\includegraphics{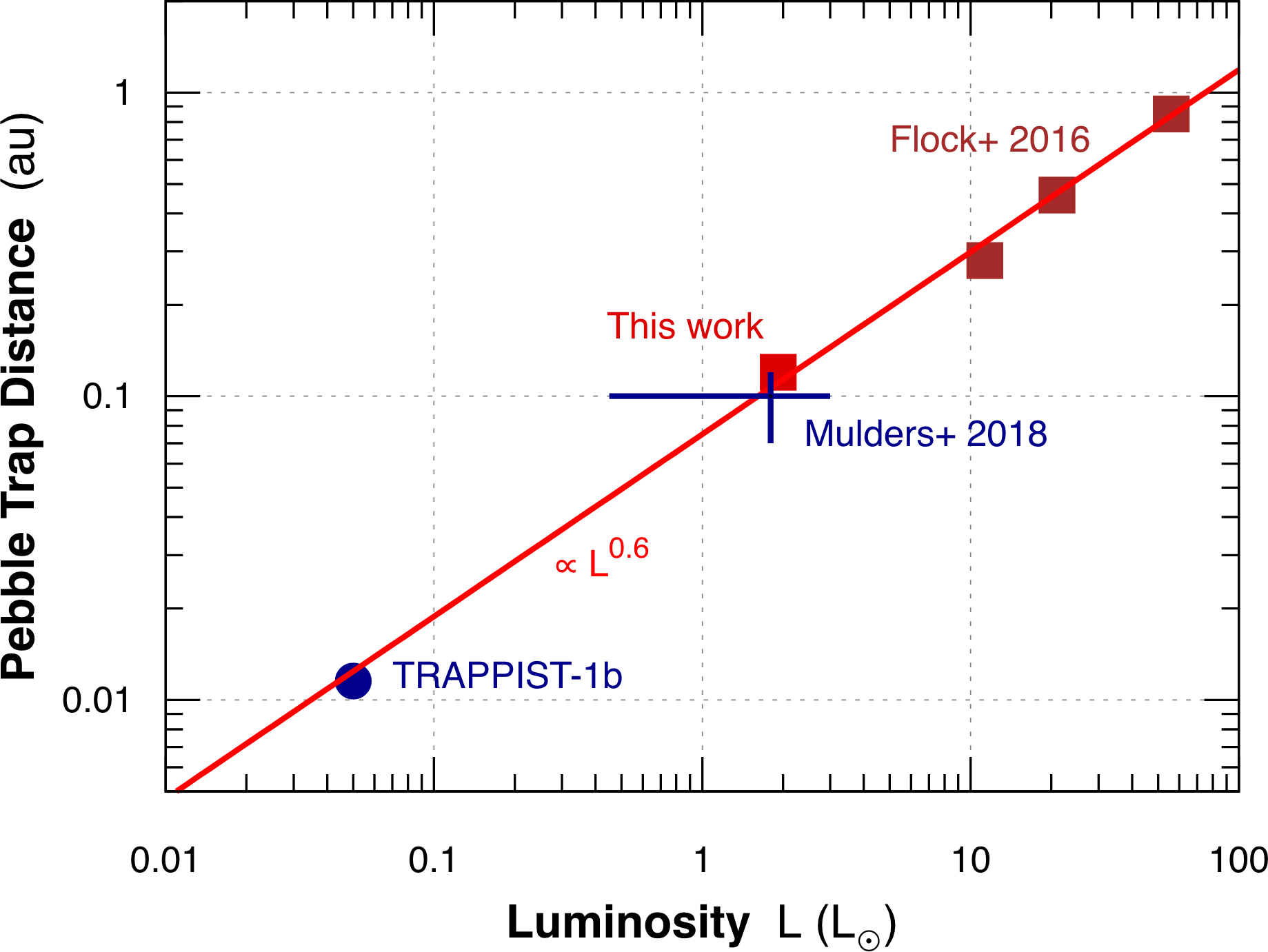}}
  \caption{Relation between the pebble trap {distance and} the luminosity of
    the star. {The leftmost red cross is from model \texttt{MREF} and the rest are from our
    previous work \citep{flo16}.}} 
\label{fig:lum}
\end{figure}

{As with the silicate} sublimation {front}, the snow line is
determined by the {stellar} luminosity. \citet{bit15} show that the snow line can
lie inside  1 au at {accretion rates similar to those we consider}. The snow line is significantly farther from the star in our models for
  two reasons.  First, the stellar luminosity is about twice as great.
  Second, the opacity for starlight exceeds that for the disk's own thermal
  emission, where \citet{bit15} had the reverse.  Absorbing efficiently and re-emitting inefficiently makes the dust hotter at a given starlight flux, pushing the snow line outward.

\section{Discussion}
In this section we discuss the effects of viscous heating, the magnetic
driven wind, and the stellar magnetic field on our model. We
compare our results to previously found migration maps and other planet formation
models. Finally, we report on the effect that eccentricity and resonant
chains might have.

\subsection{Viscously heated or passive disk}
We have  focused here on irradiated passive disks. Our similar radiation hydrodynamical models of disks around intermediate-mass young stars show little impact from accretion heating \citep{flo16}.
{However for T
  Tauri disks accretion is more important since the ratio
  of accretion flux to irradiation flux at the sublimation front varies roughly inversely with stellar
  luminosity. To examine whether accretion heating is important, in
  Fig.~\ref{fig:visc} we compare the passive irradiated disk with one heated
  only by accretion. Following \citet{hub90} the accretion flux
radiated from the disk surface is

\begin{equation}
F_{acc}=\frac{3}{8 \pi} \dot{M} \Omega^2 \left ( 1 - \sqrt{ \frac{R_0}{R}}
\right )
\end{equation}
using the corotation radius $R_0$. With the accretion power deposited in the
interior, the midplane temperature is 

\begin{equation}
T_{mid} = \left ( \frac{\tau_{Z} F_{acc}}{\sigma_b} \right )^{1/4},
\label{eq:acc_heat}
\end{equation}
{for vertical Rosseland mean optical depth $\tau_{Z}$. The midplane
  accretion temperature for the reference model's surface density profile is
  the blue dashed line in Fig.~\ref{fig:visc}.} 
\begin{figure}
  \resizebox{\hsize}{!}{\includegraphics{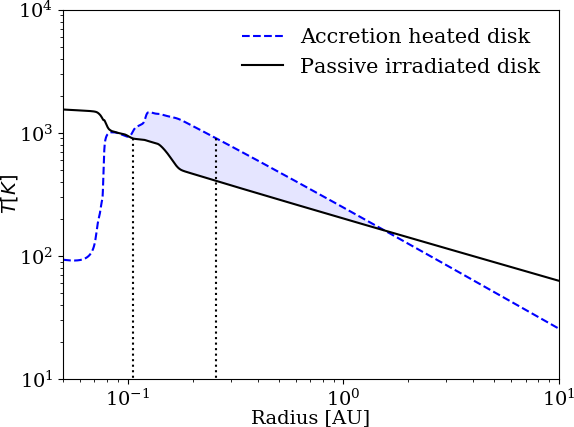}}
  \caption{{Profile of the midplane temperature in model \texttt{MREF}
      (black line) and in the model heated only by accretion using the classical
      $\alpha$ prescription (blue dashed line). The dotted vertical lines indicate the edges of the  dead zone  in MREF (left) and the $\alpha$ model (right).}} 
\label{fig:visc}
\end{figure}
The accretion heating shifts the edge of the dead zone
 from 0.1 au to 0.26 au. {However, in the passive model (solid
  line in Fig.~\ref{fig:visc}) the annulus between these two radii is cool enough to
  remain MRI inactive. Furthermore, the turbulent surface layer or base of the
  magnetized wind, where the accretion power is deposited when the interior is
  inactive, has an optical depth
  comparable to or less than unity \citep{bai09,bet17,mor19}.  According to
  Eq.~\ref{eq:acc_heat} the accretion temperature at the base of the  active layer 
  is well below the irradiation temperature, so accretion heating is
  negligible.
  Thus, the accretion flow in the passive irradiated model does not
  significantly alter its temperature. Furthermore, for the purpose of Fig.~\ref{fig:visc} we
  ignore the dust sublimation;  by speeding up cooling sublimation would act as a thermostat
  and limit the temperature to $T_s$, again bringing the temperatures close to
  those in the passive model. We also note that the accretion-heated disk is
  much colder than the passive disk inside ~0.08 au, due to the low gas
  opacity.  If both irradiation and accretion were included, the temperature
  here would be similar to the irradiation-only case.}}


In summary, the passive irradiated models accurately represent the thermal
structure. However, accretion heating is important between the inner rim and
the edge of the dead-zone {for higher mass accretion rates $\rm \dot{M} \ge 10^{-8}$, pushing the
  edge of the dead zone  and the planetary trap farther out. However, even in this case, we still expect that the
  first planetary core to form will be dry, since the dead-zone pebble trap
  always occurs at high temperatures of around 800K to 1000K.}

\subsection{Torque map comparison}
One noticeable difference from previous models is the contribution of the
entropy-related corotation torque. In our models this effect is slight, giving rise only to a small outward migration region that sits outside   the 
pebble trap in Fig.~\ref{fig:torq}.
In previous studies {for T Tauri systems} that include {a higher} viscous heating \citep{col14,bit15}, the
outward migration zone extends to 5 au and beyond because of
the steeper temperature gradient obtained in viscously heated disks. 
Following the previous discussion, we expect that the contribution 
from the entropy-related corotation torque in the dead zone should remain 
small in the absence of another source of heating.

\subsection{Gap-opening mass}
{Equation~\ref{eq:gap} yields} traditional gap-opening mass appropriate to viscous
disks. However, if $\alpha_{DZ}$ {takes the form of an external torque
  applied by magneto-centrifugally launching a wind from the disk surface, its
  character may be quite unlike a viscosity, influencing} the gap-opening mass. In this case, where the disk midplane is laminar, the
gap-opening mass is much lower \citep{raf02,don11,mcn19}, especially for the
low values of $H/R$ in our models. 

\subsection{Exact shape and position of the inner rim}
{When the sublimation front is highly unresolved, the starlight is absorbed
  in the near face of the first dust-containing cell it meets, yet the heat is
  deposited uniformly through that cell, including its unlit interior and back
  side. Therefore the cell's front side is cooler than it should be.  Yet the
  front side is what we see in the infrared if the cell is optically thick.
  Thus, the disk's thermal radiation appears at too long a wavelength and the
  front's shape and position may be computed incorrectly.  To
  resolve this issue, we developed a method that reduces the dust abundance
  smoothly across the sublimation front so that the transition from optically thin to
  thick is spatially resolved \citep{flo16}. This approach may become
  unnecessary in future when multiple grain compositions and sizes are
  included. A mix of species, each with its own sublimation temperature,
  naturally blurs the rim \citep{kam09}.}

{The rim's location suffers from uncertainty because anything near the
  star that is opaque enough allows dust to survive within its shadow.}
Our models, which assume a very low gas opacity, should
therefore be seen as producing an outermost limit with respect to the radial position of the
rim. We note that the exact location of the pebble and planetary traps are
influenced by the rim location and the location of the ionization
transition. Furthermore, we expect the small grains in the dead zone to deplete quickly.
\citet{ued18} show  that small grains {grow efficiently} in the
dead zone, due to {its weak turbulence, forming pebbles that drift inward quickly.}

One set of parameters we have not varied is the stellar 
temperature, radius, and mass. These parameters affect the position of 
the inner rim and other features in the structure of the disk, and thus should 
be investigated in the future \citep{mul15}.

\subsection{Influence of stellar magnetic field on the rim shape}
{The star's magnetic field could potentially disturb the disk near the
  sublimation front \citep{rom12}.} {To determine whether our neglect of
  the stellar field is valid, we consider a dipole with strength of 1 kilogauss at
  the stellar surface \citep{john07}.  The field strength in the midplane at
  0.1 au is 1.8 gauss.} For our reference model \texttt{MREF}, the thermal pressure at the midplane in
the same location is $\rm P_{th}=240 dyn \, cm^{-2}$. {The plasma beta is thus 1900.}  {If it penetrated the disk's plasma, a field of this strength} would be important for driving MRI turbulence.
Although we would not expect a very strong effect on the rim profile or
height, which we investigated in our previous work \citep{flo17}, we
note that the $\alpha_{MRI}=0.1$ value used in model \texttt{M2} might be more
realistic under these conditions.

\subsection{Comparison with other planet formation models}
Because our disk models produce both a pebble trap and a planet
trap, our results in principle support both an in situ planet formation 
model based on the drift and concentration of pebbles which then form planets
\citep{bol14,cha14} and a model based on the formation of planets farther from 
the star, which then migrate inward until they are halted at the planet trap
\citep{izi17,izi19}. The {migration model appears} most able to explain the fact that some systems,  such as Trappist I
which contains seven terrestial planets around an M star \citep{gil16,gil17}, are composed of 
planets in chains of mean motion resonances.
In this model the planets migrate toward the inner edge of
the protoplanetary disk until they are parked in resonant chains.
Recent work suggests that many of these resonant chains can become
unstable after the gas disk dissipates, explaining why most of the systems
discovered by the Kepler mission are not in resonance \citep{izi17,izi19,lam19}. 

Without an inner disk edge and planet trap, migration would drive all planets into 
the central star, and resonant chains would not exist because planets could
not pile up. \citet{orm17} proposed {that planetary cores form} at the snow line and then migrate inward while continuing to grow
via pebble accretion. It is noteworthy that the innermost Trappist-1 planet is located at 
around 0.01 au, which is roughly consistent with the
inner edge of the dust disk if the luminosity was at least one
order of magnitude higher than it is currently, as is expected for young,
{very low-mass stars} \citep{lau93}. 

\subsection{Eccentricity effect on the torque}

We assume an {orbital eccentricity of zero} when calculating the  torque map.
However, planets undergoing type I migration may have a small eccentricity value, 
for example due to being embedded in a turbulent region of the disk \citep{lau04} or 
because multiple planets migrate together with a small separation. 
In such a case the outward migration zone would shrink since the corotation
torque becomes weaker \citep{fen14,col16}. This could shift the planet
traps even closer to the star, depending on the details of the density profile near
the ionization transition zone. {The shift is in the right direction to} improve the fit to the occurrence rate 
of Kepler planets.
In addition, if the turbulence is particularly vigorous, the outward migration zone
located around 0.08 au may shrink, or disappear altogether if the eccentricity
{exceeds} the {disk aspect ratio of about 0.02}. This will need to be examined
in the future using radiation-MHD  simulations of planets embedded in magnetized
disks.

%
\subsection{Effect of resonant chains}
The innermost planet in a multiplanetary system could also be pushed nearer to the star by torques from the outer planets in the resonant chain \citep{car18,izi19,car19}. This might improve the agreement between the innermost planet position
and the Kepler {results of} \citet{mul18}. 
{Furthermore,} multiplanetary systems {tend} to become unstable, 
leading to scattering and collisions among the planets once the disk has dispersed,
which could explain the occurrence of super-Earths interior to any planet trap position 
\citep{izi17,izi19}.
\section{Conclusion}
We developed radiation hydrostatic models to describe the thermal and
density structure of protoplanetary disks around T Tauri stars. The
models are 2D and  axisymmetric, and include stellar irradiation, dust and gas opacities, dust sublimation, and condensation. {Magnetically driven} accretion is modeled by means of a
temperature-dependent kinematic viscosity. This dependence is chosen to capture the onset of magneto-rotational
turbulence at temperatures around 900~K. The models are inflow-equilibrium solutions with radially constant mass accretion rates. {All our models are for} a typical young Sun-like star with parameters of $T_*=4300$K, $R_*=2.6 \,R_\sun$, and 
$M_*= 1.0 \, M_\sun$. {We examine disk mass accretion rates from} $\rm \dot{M}=10^{-9} \, to \, 10^{-8} M_\sun/yr$, and dust-to-gas mass ratios between
$10^{-4}$ and $10^{-3}$, {considering grains smaller than} 10 $\rm \mu m$ that are responsible for {absorbing} incoming stellar radiation
and {reemitting it in the thermal infrared}.

The computational domain {spans} the dust sublimation {front}
and the water snow line. The resulting disk structure {is approximately a scaled-down version of the Herbig disk models obtained by \citet{flo16}:}
\begin{itemize}
\item {In} our reference model the {dust sublimation front} occurs at around 0.08 au {in the midplane}, and {curves} out to around 0.15 au at {an elevation angle} of
  $Z/R=0.08$.

\item {The next feature outward after the sublimation front is a steep increase in density} due to the drop in the ionization level and the 
corresponding drop in the accretion stress. Here   at 0.13 au we find a robust {local pressure maximum capable of trapping pebble-sized particles.}

\item {Orbital migration torques calculated for embedded super-Earths indicate that type I migration halts at around 0.13 au.} The planet migration trap {lies close to the peak (from the Kepler survey) in the orbital distribution of the innermost planets in multiplanetary} systems.

\item The {orbital period at the pebble trap in our reference model is 17 days. In a model with the dust abundance ten times lower,} the pebble trap is shifted inward to an orbital period of 10 days. 

\item For all of our models, the snow line is located close to 1.6 au at the midplane, reaching 2
au at two scale heights above the midplane. This closely matches the peak in 
the occurrence rate of giant planets from {radial velocity surveys}.
\end{itemize}
We point out that our {results can in principle support both of the leading planet formation scenarios.}
{The disk structure we  determined allows inward-drifting} pebbles to accumulate and form planets at the 
pressure maximum {a short distance outside} the dust sublimation front. {The model also allows} planet formation to occur farther out in the disk, for example at the snow line,
after which the planets migrate inward and become trapped {beneath the
  inner rim formed by the dust sublimation front}. In the future it will be important to test these 
findings using full 3D radiation-MHD simulations with embedded pebbles and planets,
so the dynamics of these bodies can be computed within the context of the complex and
turbulent flows present in the inner regions of protoplanetary disks.

\section*{Acknowledgments}
This research was carried out in part at the Jet Propulsion
Laboratory, California Institute of Technology, under a contract with the
National Aeronautics and Space Administration and with the support of the NASA
Exoplanets Research program via grant 14\-XRP14\_2-0153. M.F. has received
funding from the European Research Council (ERC) under the European Unions Horizon 2020 Research
and Innovation Programme (grant agreement No. 757957). This research was
supported in part by the National Science Foundation under Grant No. NSF
PHY-1748958. B.B. thanks the European Research Council (ERC Starting Grant
757448-PAMDORA) for their financial support. This research was supported by
STFC Consolidated grants awarded to the QMUL Astronomy Unit 2015-2018
ST/M001202/1 and 2017-2020 ST/P000592/1. Copyright 2019 California Institute
of Technology. Government sponsorship acknowledged.

\appendix

\section{Orbital migration torque maps}
{We show torque maps for all remaining models in Fig.~\ref{fig:partor}. They are directly comparable with the map for the reference model in Fig.~\ref{fig:torq}.}

\begin{figure*}
  \resizebox{0.5 \hsize}{!}{\includegraphics{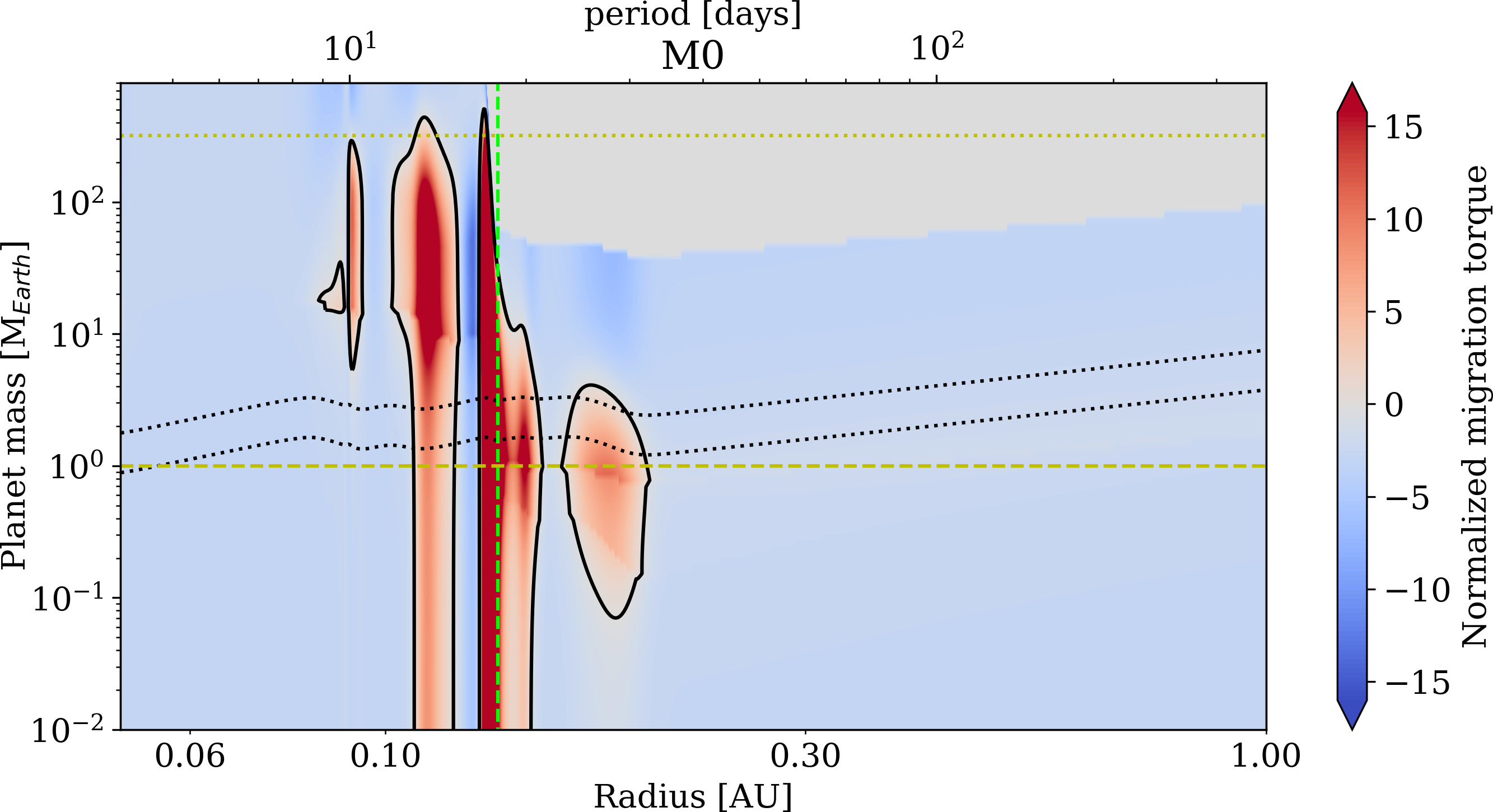}}
\resizebox{0.5 \hsize}{!}{\includegraphics{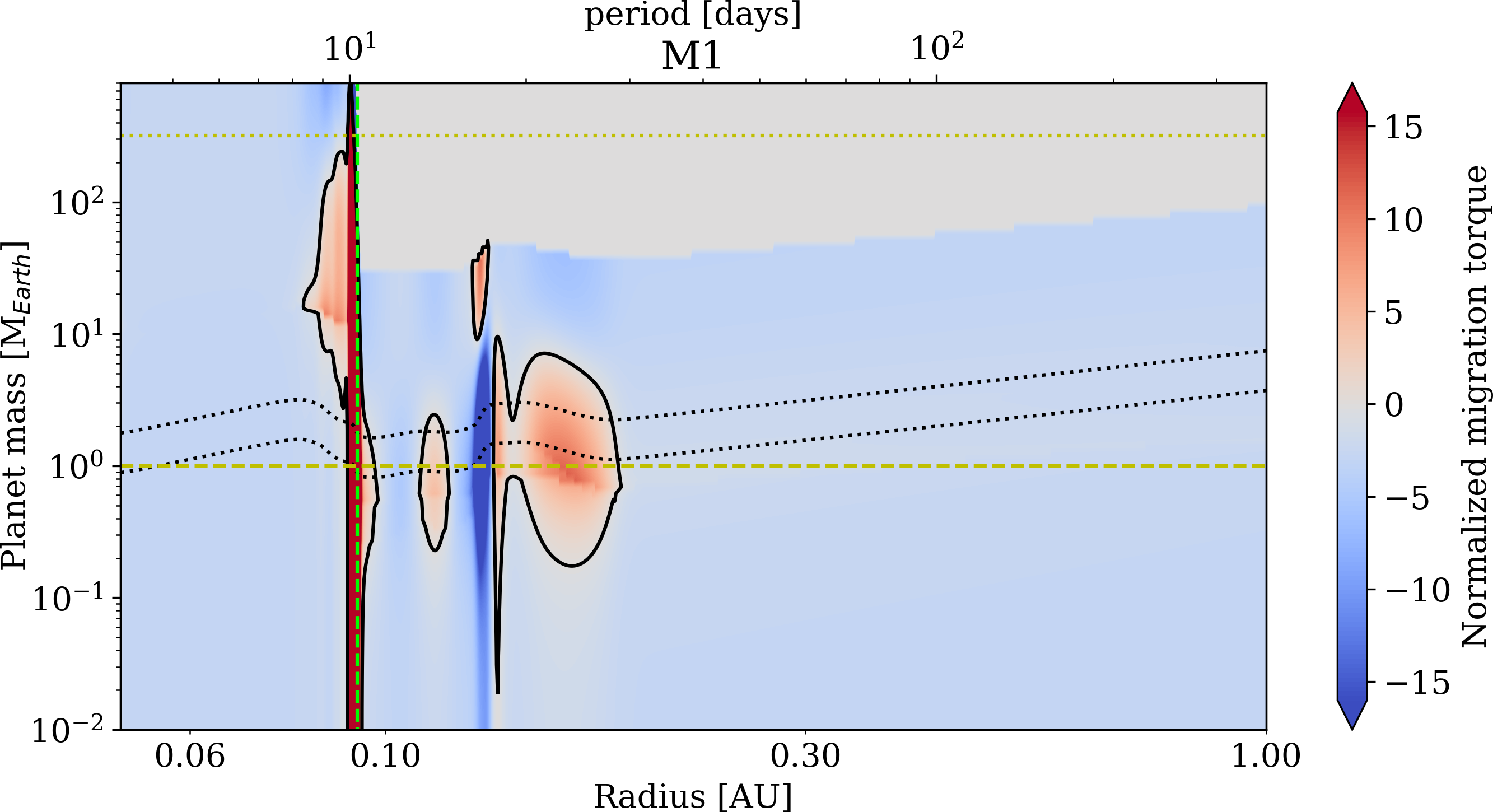}}
\vspace{2mm}\vfill{}
\resizebox{0.5 \hsize}{!}{\includegraphics{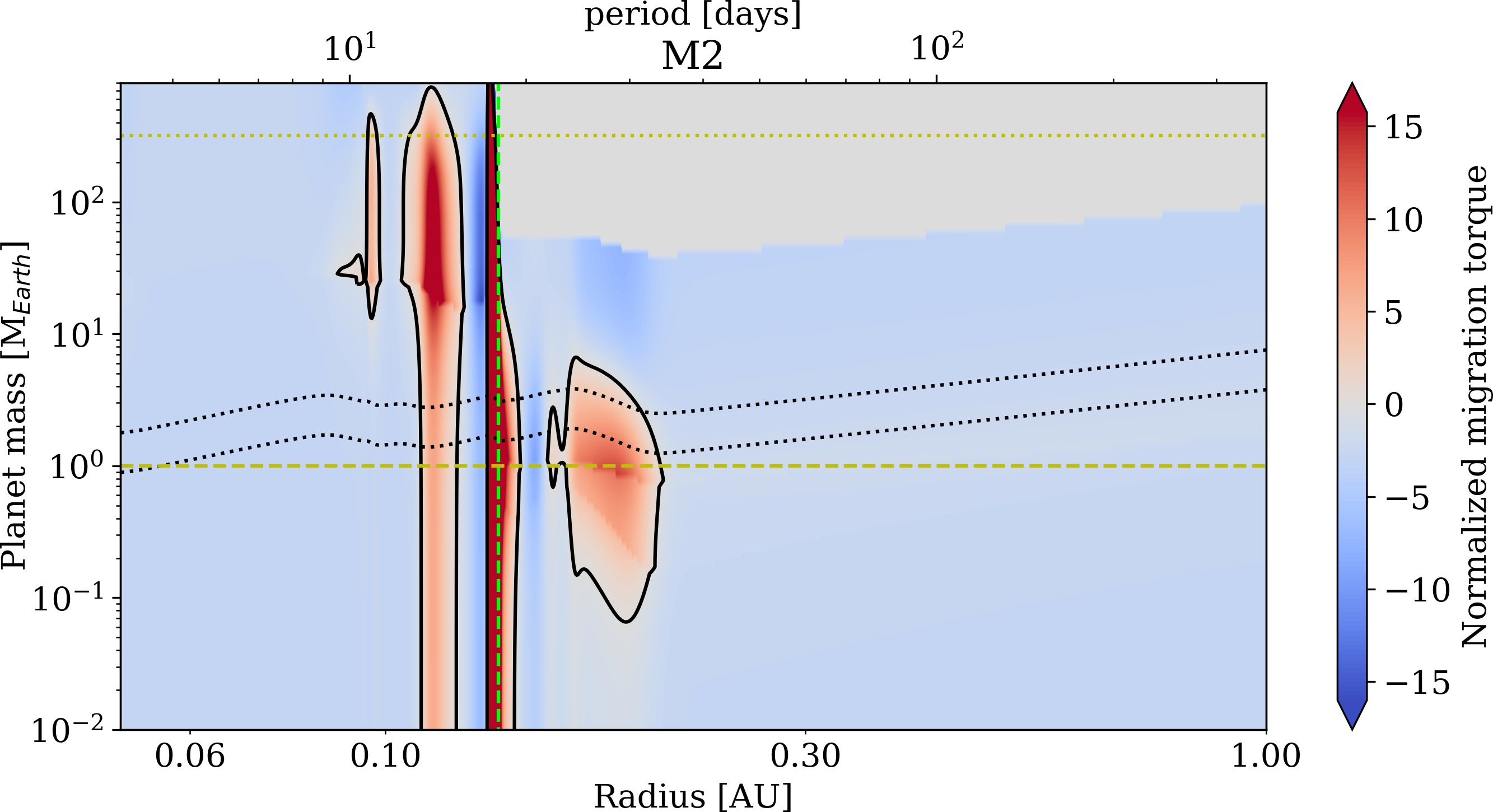}}
\resizebox{0.5 \hsize}{!}{\includegraphics{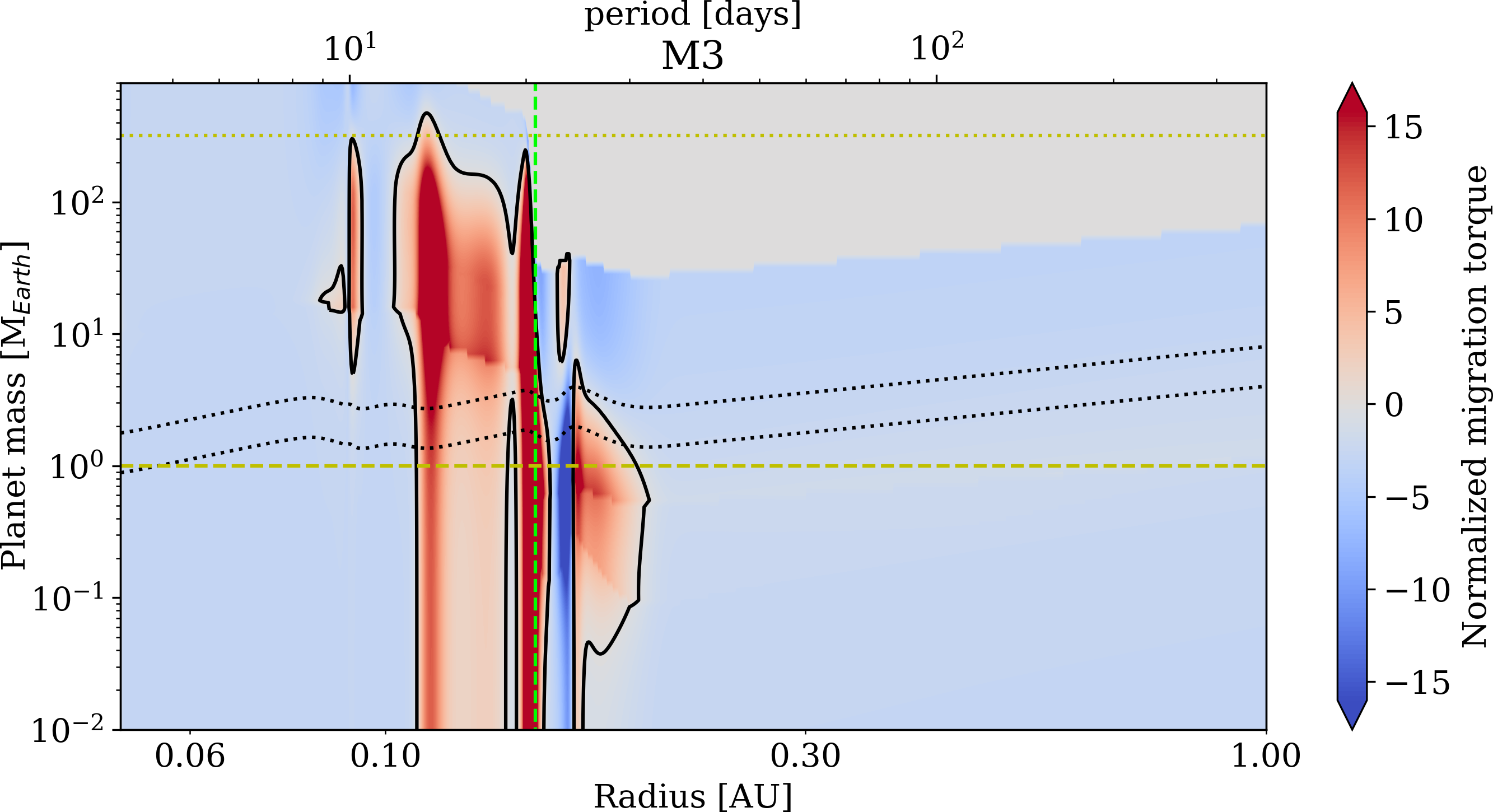}}
\vspace{2mm}\vfill{}
\resizebox{0.5 \hsize}{!}{\includegraphics{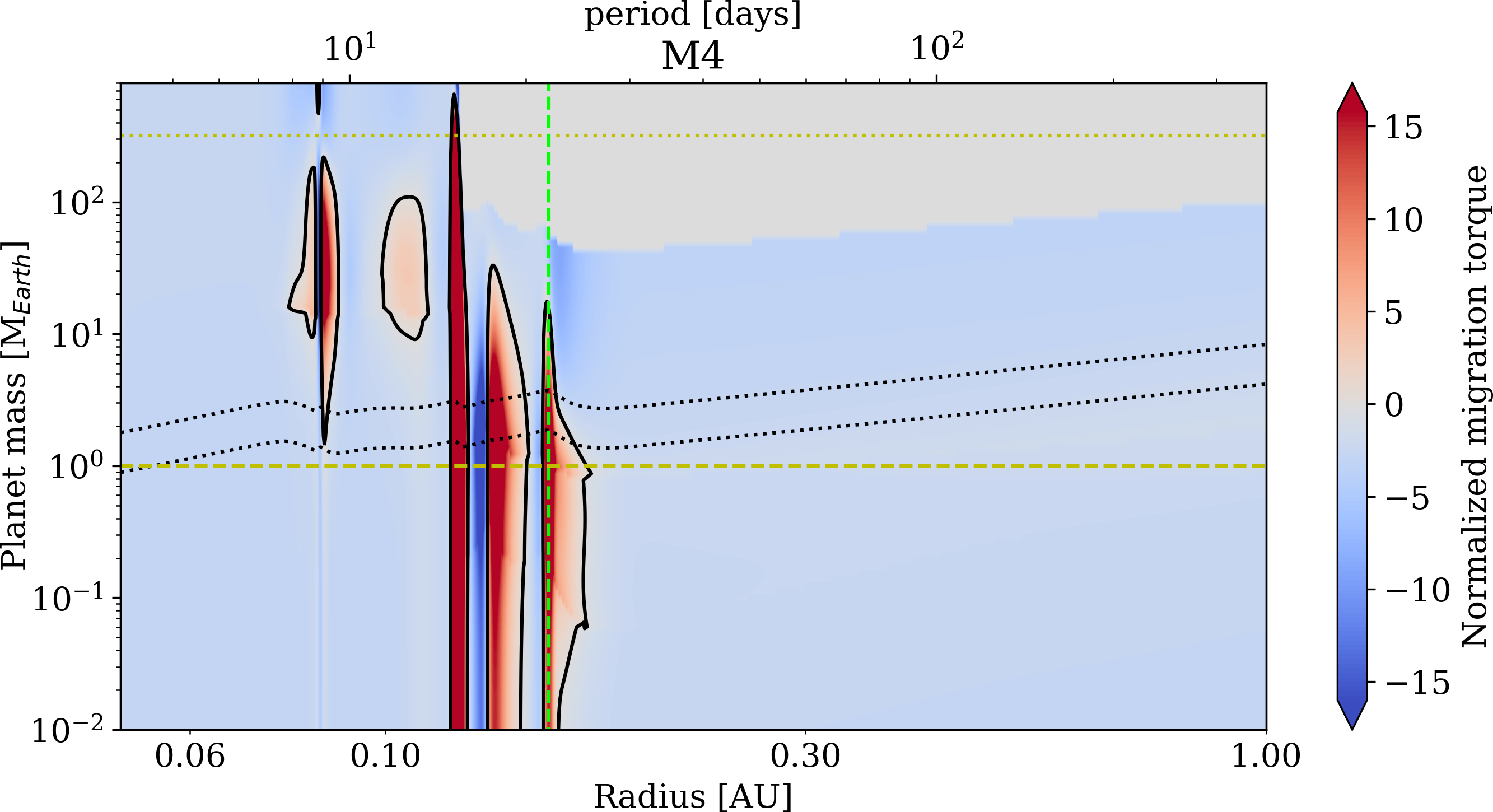}}
\resizebox{0.5 \hsize}{!}{\includegraphics{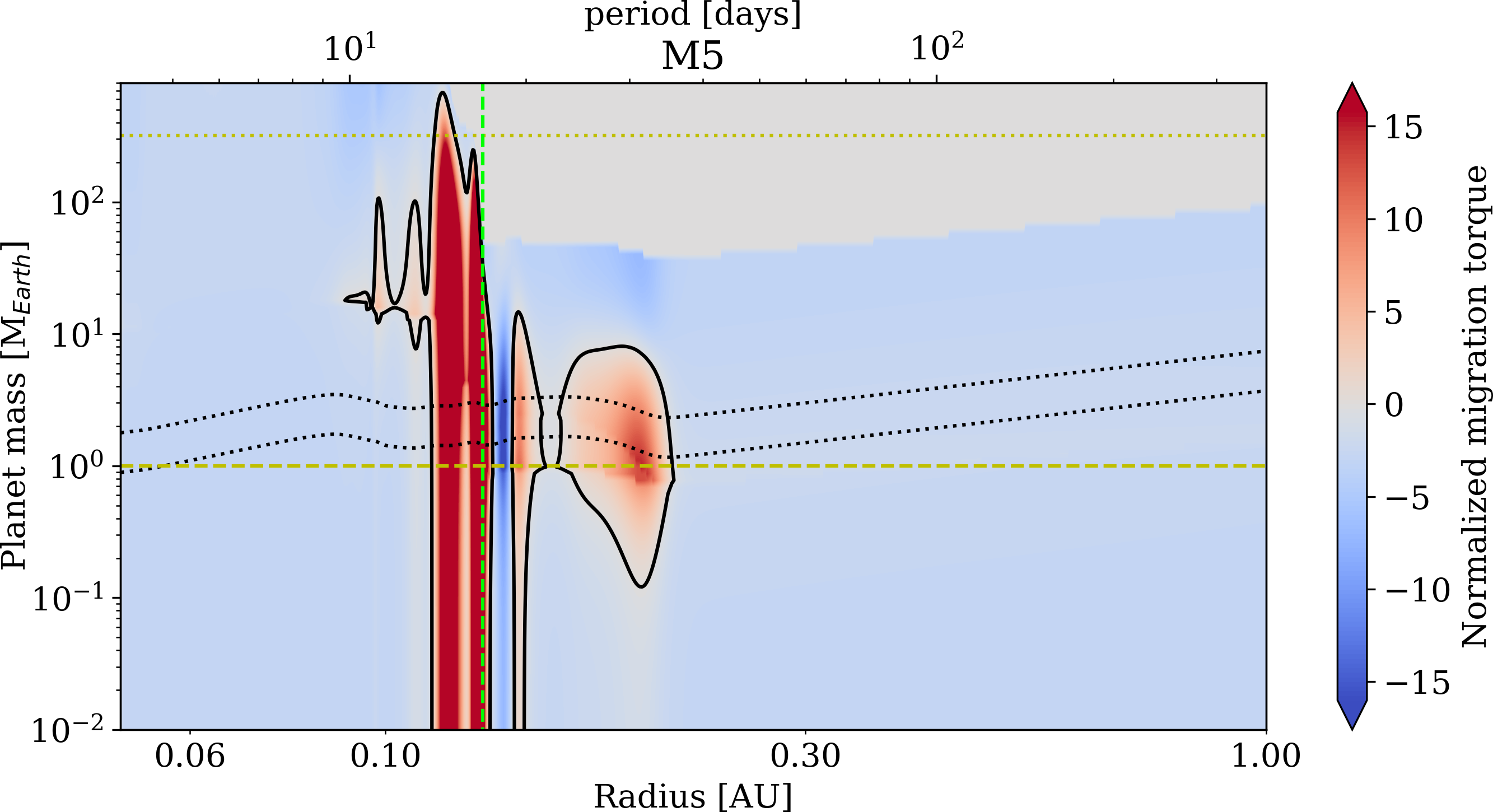}}
\vspace{2mm}\vfill{}
\resizebox{0.5 \hsize}{!}{\includegraphics{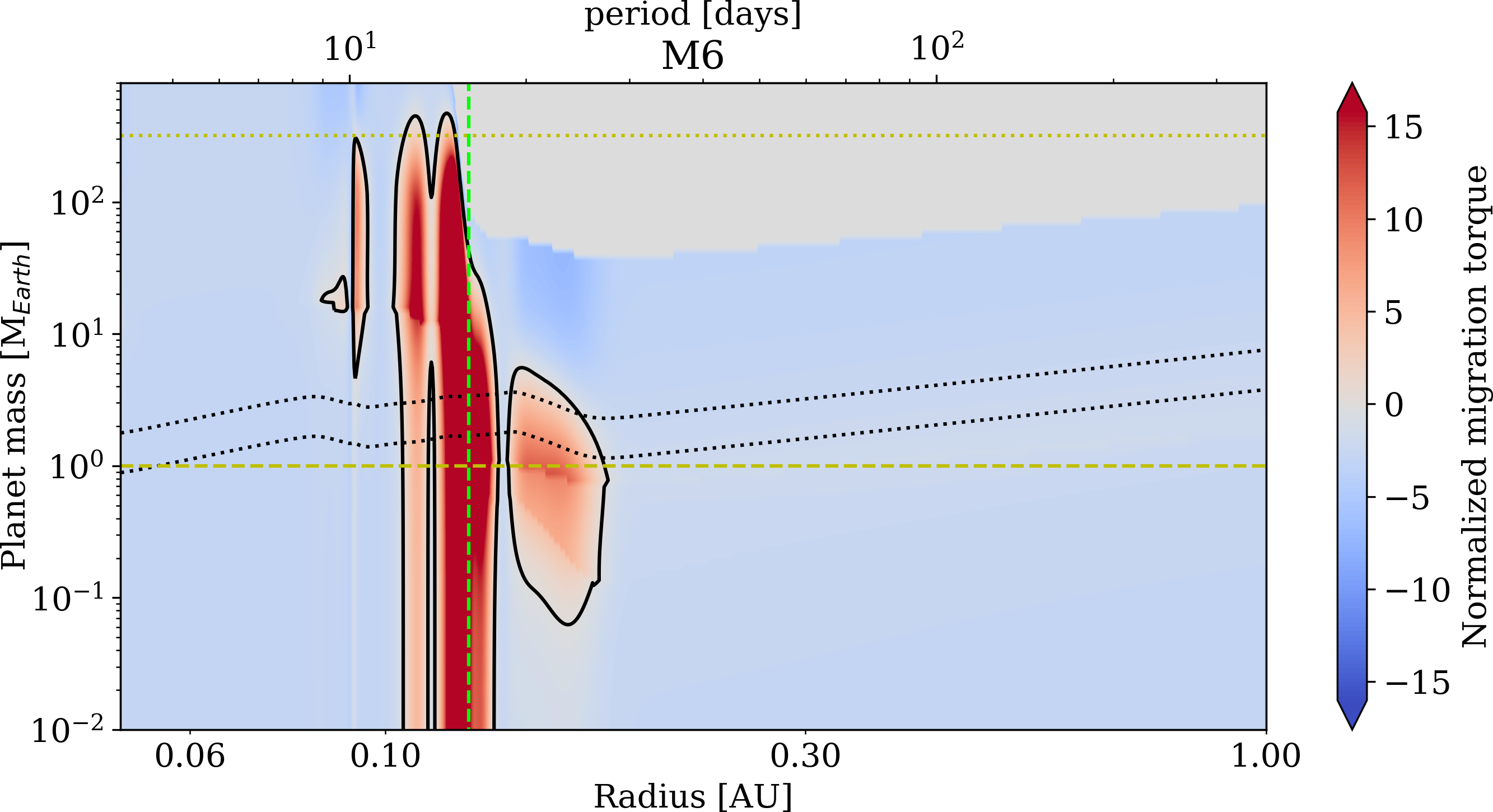}}
\resizebox{0.5 \hsize}{!}{\includegraphics{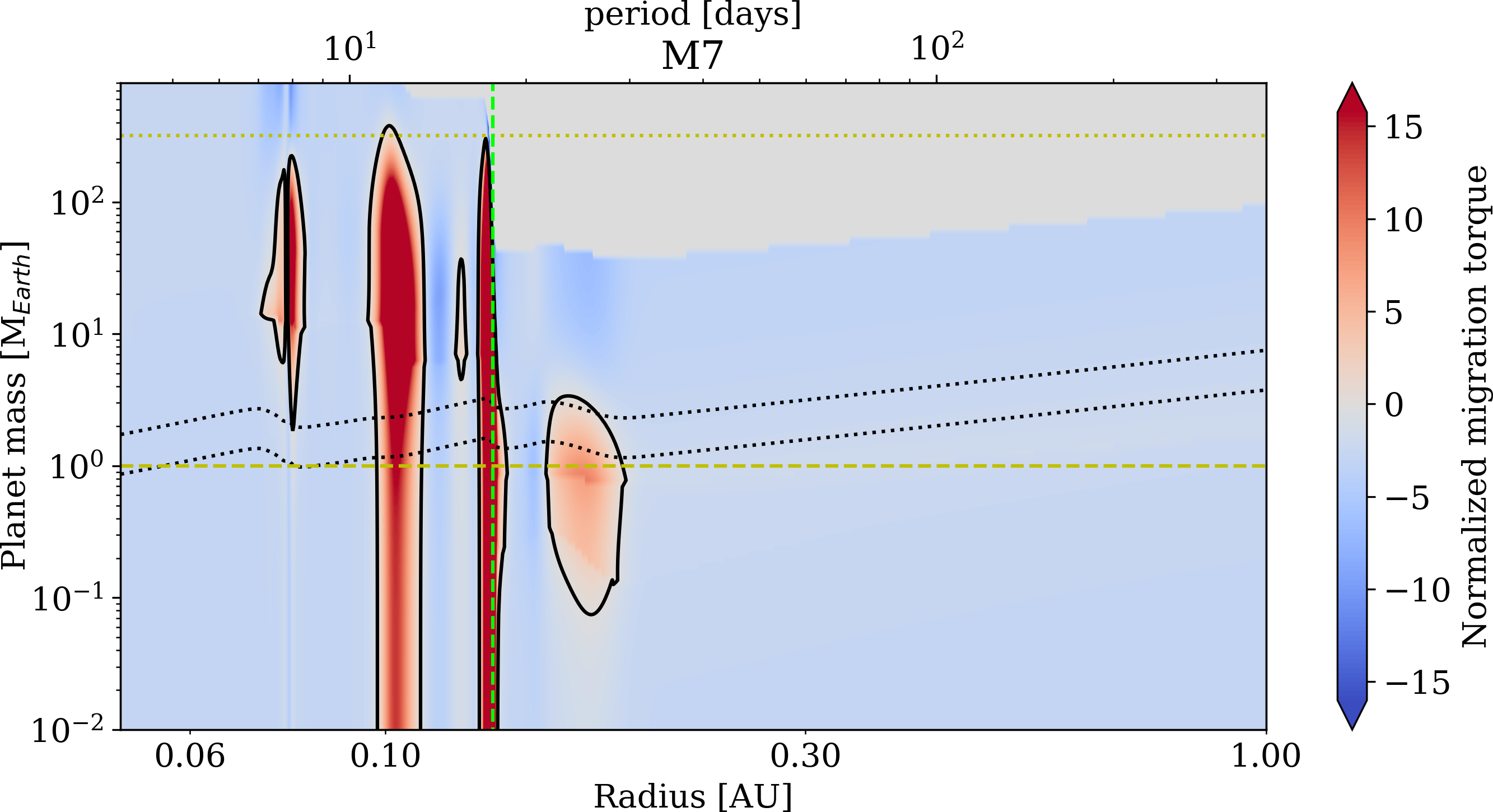}}
\caption{Torque maps {for models M0 through M7}.} 
\label{fig:partor}
\end{figure*}

\section{Opacity model}
\subsection{Gas opacity}
In our previous work \citep{flo16} we included a {simple temperature- and pressure-independent} gas opacity of $\rm \kappa_{gas}=10^{-4} cm^2
g^{-1}$. {For this new work, we make use of the more detailed} gas opacities derived
by \citet{mal14}. Figure~\ref{fig:opacd} shows the Rosseland mean gas opacities. 

{Over} the temperature and pressure regime of our model we derive a {mean} value
of about $\rm \kappa_{gas}=10^{-5} cm^2 g^{-1}$. For {simplicity} we assume the
same gas opacity also for the irradiation. Future calculations of
the detailed gas opacity including photochemistry are necessary to improve
this step. A higher gas opacity could affect the rim shape and the
temperature profile of the inner disk.
\subsection{Dust opacity}
{To compute} the dust opacity we used the tool MieX \citep{wol04}. We assumed silicate and carbon particles with radii 
between $\rm a_{min} = 0.005 \mu m$ and $\rm a_{max} = 10 \mu m$ and a grain size
distribution {having exponent} $-3.5$. {The particles are} $62.5\%$
{astrophysical silicate and $37.5\%$ graphite}. The {wavelength dependence} of the dust
opacity is plotted in Fig.~\ref{fig:opacd}.

{From the wavelength-dependent opacity we determine the Planck mean} using 
\begin{equation}
\rm \kappa_P(T) = \frac{\int_{0}^{\infty} \kappa_\nu B(\nu,T) d\nu}{\int_{0}^{\infty} B(\nu,T) d\nu},
\label{eq:planck}
\end{equation}
with the Planck function $\rm B(\nu,T)$. {The frequency integral
  corresponds to a wavelength between 0.05 $\rm \mu m$ and 2 $\rm mm$}. {For the starlight color temperature $\rm T_* = 4300\, K$, we determine  two
  significant figures $\rm \kappa_P(T_*)=1300 cm^2 g^{-1}$. For the sublimation front color temperature $\rm T_s = 1500 K$, we
determine $\rm \kappa_P(T_{s})=700 cm^2 g^{-1}$.}
\begin{figure}
  \resizebox{\hsize}{!}{\includegraphics{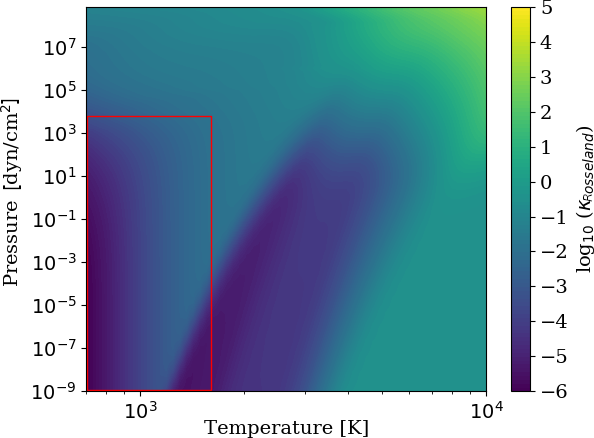}}
  \vspace{2mm}\vfill{}
  \resizebox{\hsize}{!}{\includegraphics{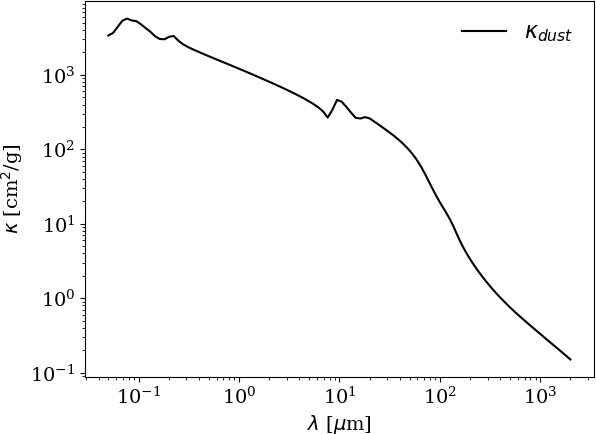}}
  \caption{{Top: Rosseland mean gas opacity vs. temperature and
      pressure} \citep{mal14}. The rectangle indicates the temperatures and
    pressures relevant for our model. {Bottom:} Total dust opacity per gram of dust over wavelength.} 
\label{fig:opacd}
\end{figure}

\bibliographystyle{aa}
\bibliography{RIM_PLANET}

\end{document}